\documentclass[%
 reprint,
 superscriptaddress,
 amsmath,amssymb,
 aps,
 prb,
]{revtex4-1}

\usepackage{graphicx}
\usepackage{dcolumn}
\usepackage{subfigure}
\usepackage{bm}
\usepackage{breakurl}
\usepackage{enumitem}

\usepackage{color}

\renewcommand{\vec}[1]{\bm{#1}}
\let\Re\relax\DeclareMathOperator{\Re}{Re}
\let\Im\relax\DeclareMathOperator{\Im}{Im}

\begin{document}


\title{Nonperturbative RG treatment of amplitude fluctuations for
  $\bm{|\varphi|^4}$ topological phase transitions}

\author{Nicol\`o Defenu}
\affiliation{Institut f\"ur Theoretische Physik, Universit\"at Heidelberg, D-69120 Heidelberg, Germany}

\author{Andrea Trombettoni}
\affiliation{CNR-IOM DEMOCRITOS Simulation Center, Via Bonomea 265, I-34136 Trieste, Italy}
\affiliation{SISSA and INFN, Sezione di Trieste, Via Bonomea 265, I-34136 Trieste, Italy}

\author{Istv\'an N\'andori}
\affiliation{MTA-DE Particle Physics Research Group, P.O.Box 51, H-4001 Debrecen, Hungary}
\affiliation{MTA Atomki, P.O.Box 51, H-4001 Debrecen, Hungary} 
\affiliation{University of Debrecen, P.O.Box 105, H-4010 Debrecen, Hungary}

\author{Tilman Enss}
\affiliation{Institut f\"ur Theoretische Physik, Universit\"at Heidelberg, D-69120 Heidelberg, Germany}

\date{\today}

\begin{abstract}
  The study of the Berezinskii-Kosterlitz-Thouless (BKT) transition in
  two-dimensional $|\varphi|^4$ models can be performed in several
  representations, and the amplitude-phase (AP) Madelung
  parametrization is a natural way to study the contribution of
  density fluctuations to non-universal quantities.  We introduce a
  new functional renormalization group scheme in AP representation
  where amplitude fluctuations are integrated first to yield an
  effective Sine-Gordon model with renormalized superfluid stiffness.
  By a mapping between the lattice $XY$ and continuum $|\varphi|^4$
  models, our method applies to both on equal footing. Our approach
  correctly reproduces the existence of a line of fixed points and of
  universal thermdynamics and it allows to estimate universal and
  non-universal quantities of the two models, finding good agreement
  with available Monte Carlo results. The presented approach is
  flexible enough to treat parameter ranges of experimental relevance.
\end{abstract}

\maketitle

\section{Introduction}
\label{sec:introduction}

The study of topological phase transitions plays a major role in
modern physics, both for the importance of having non-local order
parameters in absence of conventional spontaneous symmetry breaking
and for their occurrence in a wide variety of low-dimensional systems,
including superfluid \cite{Bishop1978} and superconducting films
\cite{Epstein1981}, two-dimensional ($2d$) superconducting arrays
\cite{Resnick1981, Martinoli2000, Fazio2001}, granular superconductors
\cite{Simanek1994}, $2d$ cold atomic systems \cite{Hadzibabic2006,
  Schweikhard2007, Murthy2015} and one-dimensional ($1d$) quantum
models \cite{Giamarchi2004}.

The standard understanding of the main properties of phase transitions
in $2d$ interacting systems is based on the role of topological
defects \cite{Nelson2002} as relevant excitations of these models.  In
$2d$ systems with continuous symmetry the unbinding of vortex
excitations drives the system out of the superfluid state above a
finite critical temperature $T_\text{BKT}$.  The mechanism for this
topological phase transition in $2d$ with continuous symmetry---in
which there is no local order parameter according to the Mermin-Wagner
(MW) theorem\cite{Mermin1966, Hohenberg1967}---was first explained by
Berezinskii, Kosterlitz and Thouless\cite{Berezinskii1972,
  Kosterlitz1973, Kosterlitz1974} and lead to the paradigm of the
BKT critical behavior\cite{Kosterlitz2016}.

The importance of the BKT mechanism can hardly be overestimated.  On
the one hand, it explained $2d$ superfluidity at finite temperature
despite the lack of off-diagonal long-range order\cite{Penrose1956},
which manifests itself in the absence of magnetization in $2d$
magnetic models such as the $XY$ model\cite{Mermin1966} and in a
vanishing condensate fraction at finite temperature in $2d$ bosonic
models\cite{Stringari1995}. Nevertheless, because of the power-law
decay of correlation functions in the low-temperature phase
\cite{Berezinskii1972} one can still have superfluid/superconducting
behavior\cite{Nelson1977}.  The physical consequences have been
studied in very different $2d$ systems, with applications ranging from
soft matter\cite{Nelson2002} and magnetic systems\cite{Bramwell1993}
to layered and high-$T_{c}$ superconductors\cite{Tinkham1996}, where
the strong anisotropy\cite{Lawrence1971} may induce BKT
behavior\cite{Tinkham1996, Benfatto2007}; for an overview of the
relevant literature we refer the reader to the recent review
\cite{Kosterlitz2016}. At the same time, despite extensive work the
effects of disorder, spatial anisotropy and more complex or long-range
interactions in real systems require the development of advanced
theoretical tools to extend our understanding of BKT topological phase
transitions to these cases.

On the other hand, $1d$ quantum systems at zero temperature can be
mapped via the quantum-to-classical correspondence to $2d$ classical
models at finite temperature and share the same universal properties.
This motivated extensive study of BKT properties in $1+1$ dimensional
models and field theories, in particular the sine-Gordon (SG)
model\cite{Coleman1975, Rajaraman1987}.  The $XY$ model can be linked
to SG theory in two steps: first, the Villain
approximation\cite{Villain1975} to the $XY$ model preserves the
periodicity of the phase variable but approximates the cosine angular
dependence with a harmonic one, and can be mapped exactly onto the
$2d$ Coulomb gas\cite{Jose1977, Minnhagen1987, Gulacsi1998}.  It was
shown rigorously\cite{Frohlich1982} that both the Coulomb gas and the
Villain model exhibit a BKT transition.  In a second step, by
  neglecting irrelevant higher vorticities/charges, the Coulomb gas is
  mapped onto the (single-frequency) 
  SG model, which also exhibits a BKT transition
\cite{Coleman1975, Amit1980, Nandori2001}.  In the Villain model,
vortex and spin-wave degrees of freedom are decoupled, unlike in the
$XY$ model \cite{LeBellac1991}.  Note that in general a strong
spin-vortex coupling can destroy the BKT transition.

The BKT scenario clearly applies when one can define local phases and
explicitly detect and study vortices, such as in the $2d$ $XY$ model
on a lattice, but it also applies in cases where it is difficult to
define the phase or detect vortices.  In all cases, the BKT transition
separates a low-temperature phase with power-law decaying correlations
from a high-temperature phase with exponentially decaying
correlations, without an explicit need to monitor vortex
configurations\cite{Hadzibabic2006}.  Even then, one cannot
  disregard the periodic (compact) nature of the phase variables,
  which is necessary to obtain the BKT transition and is correctly
  taken into account in the SG model.  The subtlety of the compact
  phase is the reason why the results from the SG and $|\varphi|^4$
models are not easily related: the SG model provides an excellent
description of the RG flow near the critical point and is the natural
formalism to include the compact phase, but the inclusion of
fluctuations apart from vortices is not straightforward.  In contrast,
within the $O(2)$ symmetric $|\varphi|^4$ theory one readily
incorporates amplitude fluctuations but it is difficult to access the properties related
to the periodicity of the phase and to recover the BKT transition in
the thermodynamic limit (see below).

In this work we discuss the role of the phase variable in
$|\varphi|^4$ theory and show that the amplitude-phase (AP) Madelung
representation of the field $\varphi=\sqrt{\rho}\, e^{i\theta}$ leads
to a consistent and efficient treatment combining the advantages of
the SG and $|\varphi|^4$ approaches.  With this tool we can treat on
equal footing both the $XY$ lattice model, where amplitude is fixed by
construction, and the $|\varphi|^4$ model, for which we show without
{\em a priori} assumptions that amplitude (density) fluctuations are
gapped at the critical point, at least at leading order in the
truncation.  For this purpose we employ an exact mapping from the $XY$
model to an appropriate $|\varphi|^4$ theory.  Our approach with a
periodic phase variable recovers the universal properties of the BKT
transition including the line of fixed points, essential scaling and
the equation of state in the fluctuation regime; this would be lost
without phase periodicity.  More importantly, we can also study the
contribution of amplitude and longitudinal spin fluctuations to
non-universal quantities such as the critical temperature, which is
useful for BKT studies of $2d$ superconductors\cite{Schneider2000,
  Benfatto2012} and other materials.

We perform our study in the framework of the functional
renormalization group (FRG), which generalizes the idea of Wilson
renormalization to the full functional form of the Landau-Ginzburg
free energy. Since its introduction\cite{Wetterich1993}, FRG has been
able to recover and expand most of the traditional RG results and
provides a systematic approach for the investigation of
high-energy\cite{Berges2002}, condensed matter\cite{Kopietz2010,
  Metzner2012, Boettcher2012} and statistical
physics\cite{Delamotte2012}.  An advantage of FRG is particularly
evident when considering the universal critical exponents of $O(N)$
field theories as a function of the spatial dimension $d$ and the
field component number $N$.  The FRG approach combined with lowest
order derivative expansion \cite{Morris1997} gives numerical
  results for the anomalous dimension $\eta$ and correlation length
  exponent $\nu$, which reproduce the expected behavior in the
limiting cases $N\to \infty$, $d\to 4$ and $d\to 2$ \cite{Codello2013,
  Codello2015}; also $O(N)$ models with long-range interaction have
been studied \cite{Defenu2015, Defenu2016}.

Since several works already addressed the BKT transition using FRG
\cite{Grater1995, Gersdorff2001, Krahl2007, Nagy2009, Machado2010,
  Jakubczyk2014, Jakubczyk2017}, we think it is useful to explain here
in detail our motivation to study $2d$ systems in an FRG framework
using the AP parameterization.  FRG reproduces for $d \to 2$ the exact
behavior required by the MW theorem \cite{Mermin1966, Hohenberg1967}.
Moreover, it is possible to recover the MW theorem already at the
lowest order of the derivative expansion, \emph{i.e.}, in the local
potential approximation \cite{Defenu2015-2}. The compatibility of FRG
results with the MW theorem also leads in the $N>2$ case to an exact
agreement of numerical critical exponents with the lowest order
$4-\varepsilon$ \cite{LeGuillou1980} and $2+\tilde{\varepsilon}$
expansion \cite{Brezin1976} for the $O(N)$ nonlinear $\sigma$ models.
Furthermore, for the anomalous dimension $\eta$ in general $d$ one
finds $\eta \to 0$ for $d > 2$ and $N \ge 2$ in the limit $d \to 2$
\cite{Codello2013, Codello2015}.  However, in the BKT case $d=2$ and
$N=2$, the application of FRG is much less straightforward.

The field theoretical and FRG approaches to the $d=2$, $N=2$ case in
general use a two-component, complex $|\varphi|^4$ theory in the
continuum. The field $\varphi$ entering the partition function can be
parametrized in the following ways:
\begin{itemize}
\item[(i)] the field and its complex conjugate, $\varphi$ and $\varphi^\ast$;
\item[(ii)] the real and imaginary parts of $\varphi$, \emph{i.e.},
  $\Re{\varphi}$ and $\Im{\varphi}$;
\item[(iii)] the amplitude $\rho$ and phase $\theta$ of the field
  $\varphi=\sqrt{\rho}\, e^{i\theta}$.
\end{itemize}

In the paper \cite{Grater1995} the $|\varphi|^4$ model in $d=2$ is
studied within FRG by the derivative expansion formalism using the
parametrization (i), where the phase periodicity is implicitly
implemented. Proceeding in this way, one can show that there is a line
of (pseudo)-fixed points, which is a hallmark of BKT, and $\eta$ can
be estimated in good agreement with the BKT prediction, even though it
is not possible to unambiguously locate the critical point.  Indeed,
in order to locate the critical point it is necessary to terminate the
FRG flow at a finite scale, corresponding to a reasonable (but
arbitrary) size of the system as also used in\cite{Krahl2007}. The
$\beta$ function for the interaction coupling $\lambda$ obtained in
this FRG scheme agrees with the one of the nonlinear $\sigma$ model
only at first order in the temperature $T$. This discrepancy leads to
a rather different behavior: in the loop expansion of the non-linear
$\sigma$ model the flow of the interaction $\lambda$ is trivial since
all loop contributions vanish, and the model remains always in its
low-temperature phase. On the other hand, the FRG
treatment\cite{Grater1995} gives a nontrivial flow for the $\lambda$
coupling with a line of pseudo-fixed points appearing at low
temperature and a high-temperature phase where the system renormalizes
to a symmetric state with $\lambda\to0$; this is interpreted as a hint
of BKT behavior.  However, the low-temperature pseudo-fixed points in
the FRG flow are unstable and the system is always driven to the
high-temperature state in the thermodynamic limit, in contradiction to
the BKT picture.  This instability remains also in FRG with
higher-order truncations\cite{Gersdorff2001,Jakubczyk2017}.

Parametrization (ii)\cite{Jakubczyk2014, Jakubczyk2017} has the
advantage that the transverse mode ($\Im\varphi$) alone reproduces the
BKT scenario, if one disregards the massive longitudinal
($\Re\varphi$) mode.  However, in view of more complex cases in which
the existence of the BKT transition is not \emph{a priori} known, it
is important to study also the effect of the massive longitudinal mode
(in which case the flow equations derived in parametrization (ii)
become equivalent to (i)).  It turns out that the interplay of
massless transverse modes with the massive longitudinal mode makes the
line of fixed points unstable and drives the RG flow to the high
temperature phase for all initial conditions \cite{Jakubczyk2017}.
Instead, with a temperature dependent regulator that is optimized
(fine-tuned) for each initial condition of the RG flow, a line of true
fixed points is found with very good results for the anomalous
dimension and the jump of the stiffness at $T_\text{BKT}$
\cite{Jakubczyk2014}.

In this paper we argue that FRG in the AP parametrization (iii)
overcomes possible ambiguities in the other parametrizations and
achieves two goals: first, it recovers the BKT transition in the $XY$
and $|\varphi|^4$ models without any {\em ad hoc} assumption on its
existence and validity; and second, it quantifies the effect of
amplitude fluctuations on the superfluid stiffness and nonuniversal
properties of both models.  The paper is structured as follows:
Section~\ref{sec:models} defines the models and recapitulates previous
FRG results; Section~\ref{sec:mapping} explains the mapping from the
lattice $XY$ model to the continuum $|\varphi|^4$ model so that we can
treat both on equal footing.  Section~\ref{Sec4:AP} introduces our new
FRG approach, which proceeds in two stages: first, we formulate the
FRG in the AP parametrization to integrate over amplitude and
longitudinal phase fluctuations; at the end of this flow we obtain an
effective SG model with a renormalized superfluid stiffness.
Subsequently, transverse phase (vortex) excitations in the SG model
with compact phase drive the BKT transition and yield a line of true
fixed points.  In Sec.~\ref{sec:res} we present our results for the
$|\varphi|^4$ model, where we recover the universality of
thermodynamic functions in the fluctuation regime \cite{Popov1983,
  Fisher1988, Baym1999, Prokofev2001, Prokofev2002, Pilati2008,
  Rancon2012}, as well as for the $XY$ model where we discuss the
temperature dependent renormalization of the superfluid stiffness.
Finally, we conclude in Sec.~\ref{sec:concl}.

\section{The models and discussion of previous FRG results}
\label{sec:models}

In this section we introduce the $XY$ and $|\varphi|^4$ models studied
in this work and recapitulate basic properties of the BKT phase
transition.  We then discuss previous FRG work before presenting our
results in Secs.~\ref{Sec4:AP}--\ref{sec:res}.

\subsection{The $\bm{XY}$ and $\bm{|\varphi|^4}$ models in $\bm{2d}$}
\label{sec:XY}

The Hamiltonian of the XY or plane rotor model reads
\begin{align}
\beta H_{XY}= - K \sum_{\langle ij\rangle} 
\left[\cos\left(\theta_{i}-\theta_{j}\right)-1\right]
\label{eq:XY}
\end{align}
where $K=\beta J>0$ denotes the spin coupling in units of temperature
and as usual $\beta=1/k_B T$.  The angles $\theta_i$ are defined at
the sites $i$ of a $2d$ lattice; in the following we consider a square
lattice.  The ground state is fully magnetized with all spins pointing
in the same direction, $\theta_{i}=\theta_0\,\forall\,i$, and is
infinitely degenerate.  At any $T>0$ symmetry breaking is forbidden in
$2d$ by the MW theorem.  Nevertheless, finite systems can have a
nonzero magnetization, which is used to detect the BKT transition
\cite{Bramwell1994}.  In ultracold atomic gases the counterpart of the
magnetization is the $\vec{k}=0$ component of the momentum
distribution and the central peak of the atomic density profile
sharply decreases around $T_\text{BKT}$ \cite{Trombettoni2005}.

The action for the $|\varphi|^4$ model reads
\begin{align}
\label{Eq13}
S[\varphi]=\int d^{2}x\left\{\frac{1}{2 m}\partial_{\mu}\varphi\partial_{\mu}\varphi^{*}-\mu|\varphi|^{2}+\frac{U}{2}|\varphi|^{4}\right\}.
\end{align}
Note that Eq.~\eqref{Eq13} has been written for a classical field
  $\varphi$, but in the following it will be applied also to the
  interacting boson case.  There, $\mu$ would represent the chemical
  potential, $U$ the local interaction and $m$ the boson mass.  In the
  following we will use unit mass $m=1$ but restore it when
  convenient.  We shall use units in which $\hbar=k_B=1$.
  
Continuous $O(N)$ field theories, with the action \eqref{Eq13}
corresponding to $N=2$, have been studied extensively and provide
important examples of the field theoretical treatment of phase
transitions.  The nonperturbative FRG has produced a comprehensive
picture of the universality classes of such theories for every real
dimension $d$ and number of field components $N$
\cite{Codello2013,Codello2015}.  In Section~\ref{sec:mapping} we
discuss how to map the lattice $XY$ model \eqref{eq:XY} onto the
continuum $|\varphi|^4$ theory \eqref{Eq13}.

To fix the notation and state results used later, we briefly
recapitulate basic results of the BKT universality class
\cite{Pelissetto2002} referring to the $XY$ model. A discussion of BKT
theory in the $|\varphi|^4$ model can be found, e.g., in
\cite{Prokofev2002}.  Within a spin-wave analysis of the $XY$ model
\eqref{eq:XY}, we can expand around the symmetry broken state for
small phase displacements $\theta_{i}-\theta_{j}\ll 1$, which in the
continuum limit leads to
\begin{align}
\label{Eq2}
\beta H_\text{sw}=\frac{K}{2}\int{\left(\nabla\theta\right)^{2}d^{2}x}.
\end{align}
When the phase $\theta$ is treated as periodic the latter model is
equivalent to the Villain model\cite{Villain1975}.  Neglecting the
compactness of phase variable $\theta$, one readily finds
\begin{align}
M_{L} \propto\left(\frac{a}{L}\right)^{\frac{1}{2\pi K}}, \,\,\,\,\, 
G(x) \propto\left(\frac{a\pi}{x}\right)^{\frac{1}{2\pi K}},\label{Eq4}
\end{align}
where $M_{L}$ is the magnetization of a finite system of size $L$ and
lattice spacing $a$, and $G(x)$ denotes the two-point correlation
function between two spins at distance $x$ in the thermodynamic limit
(see App.~\ref{AppendixA} for a derivation).

The magnetization $M_L$ decays as a power law of the system size, and
in the thermodynamic limit the system has no finite order parameter at
finite temperature, in agreement with the MW theorem.  On the other
hand, the two-point correlation $G(x)$ displays algebraic behavior
with temperature dependent anomalous dimension
\begin{align}
\eta(T)=\frac{T}{2\pi J}.
\end{align}
This result is generally valid also at higher order in the
low-temperature expansion of the system.
 
The spin-wave analysis suggests that the ordered phase is stable at
all temperatures and the correlation functions have power-law behavior
even for small $K$ values. However, this is inconsistent with an
intuitive argument \cite{Kosterlitz1973} based on the free energy
$F=(\pi J -2T)\log\left(\frac{L}{a}\right)$ of a macroscopic vortex
configuration (see, e.g., \cite{LeBellac1991}).  Accordingly, vortex
configurations of the spin should become favorable for temperatures
larger than
\begin{align}
\label{Eq7}
T_\text{BKT} \approx \frac{\pi J}{2}.
\end{align}
For $T>T_\text{BKT}$, one expects vortex excitations to proliferate
and destroy the long-range order found in the spin-wave
analysis. Monte Carlo simulations have established
$T_\text{BKT} \simeq 0.893 J$ \cite{Gupta1988, Schultka1994,
  Olsson1995, Hasenbusch2005, Komura2012}.  A review of the critical
properties of the Villain model is provided in \cite{Kleinert1989},
and for comparison its critical temperature is $\simeq 1.330 J$
\cite{Janke1993}.

The continuous field theory for the spin-wave approximation is,
however, not suited to account for vortex configurations, which are
characterized by
\begin{align}
\oint_{C}\nabla\theta\cdot d\vec\ell=2\pi m_{i}
\end{align} 
when integrating over a closed contour $C$. The single-valued complex
field $\varphi$ allows for differences in the phase field $\theta$ by
multiples of $2\pi$, and thereby imposes the condition
$m_{i}\in \mathbb{Z}$ for the winding number of the vortex
configurations.  Instead, the path integral formulation with a
single-valued field $\theta$ does not include vortex configurations.

It is possible to take exact account of the vortex configurations by
means of a dual transformation \cite{Jose1977}. One can extract the
contribution from the multivalued configuration by means of the
decomposition $\theta(x)=\theta'(x)+\tilde{\theta}(x)$, where
$\oint_{C}\nabla\theta'=0$ and
$\oint_{C}\nabla\tilde{\theta}=2\pi m_{i}\neq 0$.  Substituting this
into Eq.~\eqref{Eq2}, one can show that the vortex part of the XY
Hamiltonian in $2d$ is equivalent to a Coulomb gas
\cite{Minnhagen1987} with charges playing the role of vortices. More
precisely, it is the Villain model that can be exactly mapped onto the
Coulomb gas, and spin-wave--vortex interactions give rise to
additional contributions that can be computed.  In absence of a
magnetic field, the mapping leads to a neutral Coulomb gas with
$\sum_{i}m_{i}=0$.  The Coulomb gas formalism allows for a sensible
low temperature expansion, indeed for $T\leq T_\text{BKT}$ we expect
only singly charged vortices to be relevant and we thus include only
$m_{i}=\pm 1$ configurations.  The latter give rise to an additional
cosine potential in the spin-wave Hamiltonian, and the duality
transformation maps this to the SG model in the dual phase field
$\Phi$,
\begin{align}
\label{eq:SG}
S_{SG}[\Phi]=\int d^{2}x\left( \frac{1}{2}\partial_{\mu}
\Phi\partial_{\mu}\Phi-u \cos{\left( \beta \Phi \right)} \right) , 
\end{align}
with dimensional coupling $u$ and dimensionless SG coupling $\beta$
(not to be confused with the symbol $\beta=1/k_BT$).  Also the $XY$
model can be mapped onto the SG model \eqref{eq:SG} via the Coulomb
gas \cite{Jose1977}; this can be intuitively understood because the
compact nature of the variable $\theta$ allows only perturbations in
the form of a periodic operator.  Thus, from the RG point of view the
theory space of a periodic field $\theta$ is naturally described, at
least at lowest order, by the SG model \cite{ZinnJustin1996}. Note,
however, that the original compact phase $\theta$ is replaced by the
dual phase $\Phi$ in the SG model.  We also observe that the mapping
between the $XY$ and the SG model \cite{Jose1977, Kleinert1989,
  Benfatto2012} has the advantage of giving an explicit form for the
bare coupling of the SG model.

A key point, which we will use in the following, is that the spin-wave
Hamiltonian \eqref{Eq2} with a \emph{compact} variable $\theta$ is
equivalent to the Villain model, which, once vortices with
$|m_i|>1$ are neglected, is dual to the SG model \eqref{eq:SG} with
$\beta^2=4\pi^2 K$, which becomes critical at $\beta^2=8\pi$.

The SG model has also been studied extensively in the FRG framework,
which provides a nonperturbative generalization of the original
Kosterlitz-Thouless RG equations\cite{Nandori2001,Jentschura2006,
  Nagy2008,Nagy2009,Nandori2009,Nandori2010,Bacso2015}. In the
following, after briefly reviewing in Section \ref{sec:prev} previous
FRG work for the $O(2)$ model, we will combine the AP parametrization
of the FRG with the SG results into a comprehensive FRG treatment
including amplitude, spin-wave and vortex excitations.

\subsection{FRG results for the $\bm{O(2)}$ model in $\bm{2d}$}
\label{sec:prev}

\begin{figure}[ht!]
\includegraphics[width=0.45\textwidth]{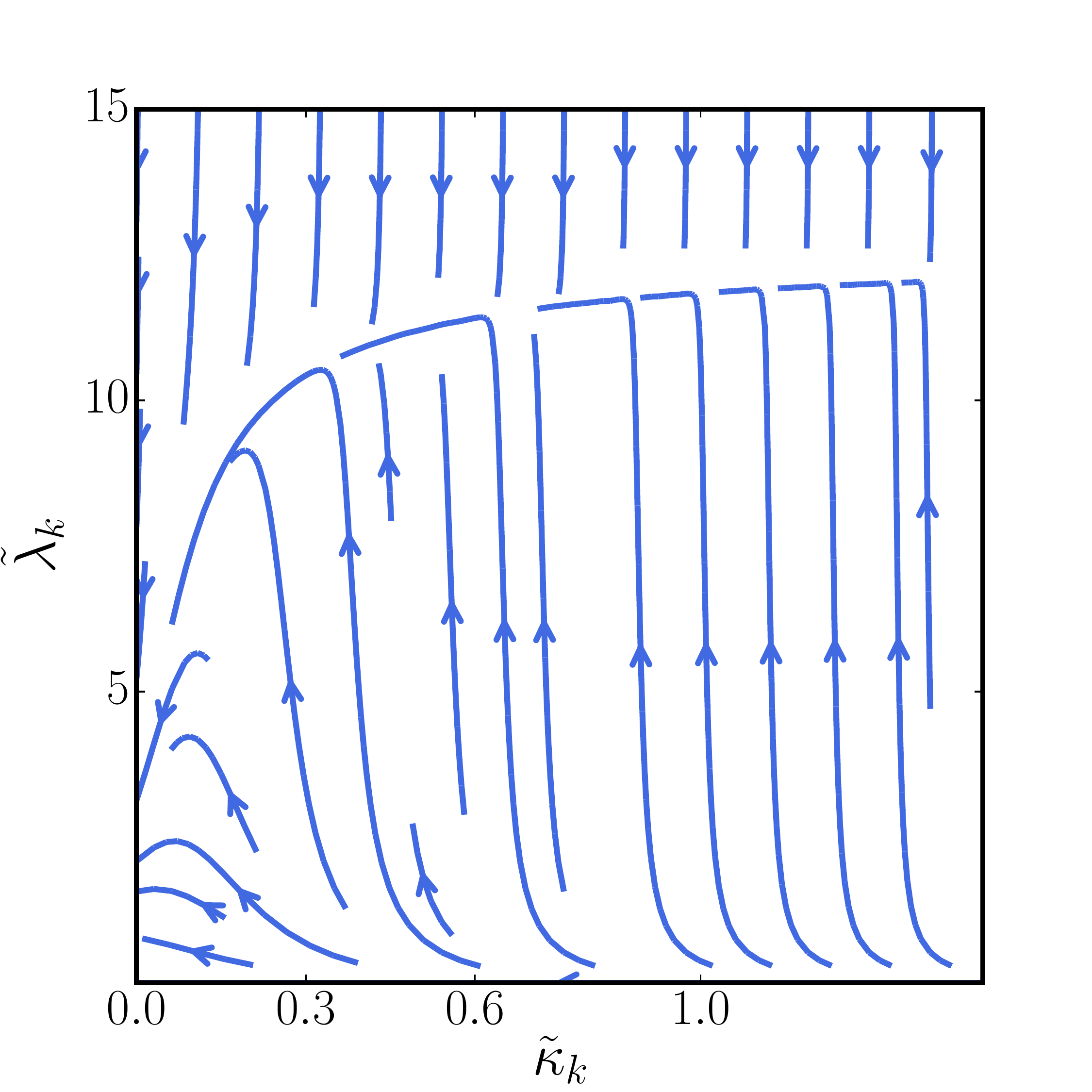}
\caption{Flow diagram of the $O(2)$ symmetric $|\varphi|^4$ theory
  with flowing effective potential \eqref{Eq14} in parametrization
  (i). The flow is first attracted toward a line of pseudo-fixed
  points at large $\tilde{\kappa}_{k}$; then the flow proceeds very
  slowly along this line toward
  $(\tilde{\kappa}_{k},\tilde{\lambda}_k)=(0,0)$, which corresponds to
  the high-temperature phase.}
\label{Fig1}
\end{figure}

In this section we review and discuss previous FRG results for the
$O(N=2)$ field theory in $d=2$. One can write the quartic potential
with $\rho=|\varphi|^2$ as
\begin{align}
\label{Eq14}
U(\rho)=\frac{\lambda_{k}}{2}(\rho-\kappa_{k})^{2}
\end{align} 
and derive FRG equations for the flow of the scale-dependent
couplings. In the LPA$'$ approximation \cite{Berges2002} one has
\begin{align}
\partial_{t}\tilde{\lambda}_{k}=(2-2\eta_{k})\tilde{\lambda}_{k}-\tilde{\lambda}_{k}^{2}\frac{\left(4-\eta_{k}\right)}{8\pi}\left(N-1+\frac{1}{(1+2\tilde{\kappa}_{k}\tilde{\lambda}_{k})^{3}}\right),
\end{align}
where $k\propto L^{-1}$ is an infrared momentum cutoff,
$\tilde{\lambda}_{k}=k^{-2}\lambda_{k}$ and
$t=-\log\left(ka\right) = 0\dotsc\infty$ is the RG ``time''. The flow
equation for $\tilde{\kappa}_{k}$ reads
\begin{align}
\partial_{t}\tilde{\kappa}_{k}=\eta_{k}\tilde{\kappa}_{k}-\frac{\left(4-\eta_{k}\right)}{16\pi}\left(N-1+\frac{1}{(1+2\tilde{\kappa}_{k}\tilde{\lambda}_{k})^{2}}\right),
\end{align}
with $\tilde{\kappa}_k=Z_{k}\kappa_k$. The anomalous dimension at scale $k$
is given by
\begin{align}
\eta_{k}=\frac{1}{\pi}\frac{\tilde{\kappa}_{k}\tilde{\lambda}_{k}^{2}}{(1+2\tilde{\kappa}_{k}\tilde{\lambda}_{k})^{2}}\label{Eq15}.
\end{align}
These flow equations for $\tilde{\lambda}_{k}$ and
$\tilde{\kappa}_{k}$ may easily be integrated numerically
\cite{Grater1995}, and the resulting phase diagram is shown in
Fig.~\ref{Fig1}.  For initial conditions with sufficiently large
$\tilde{\kappa}_{k}$ the flow is rapidly attracted to a line of
pseudo-fixed points at an almost constant value of
$\tilde{\lambda}_k$. Once this line is reached, the flow slows down
substantially and leaves the system in its symmetry broken phase for
intermediate RG times. For larger $t\to\infty$ the flow eventually
escapes the low-temperature phase and reaches the high-temperature
phase with $\tilde{\kappa}_k = 0$ at a finite time $t<\infty$.

Following \cite{Grater1995}, one can identify the unstable
pseudo-fixed line at finite $\tilde{\lambda}_k$ with the
low-temperature phase of the BKT transition.  In this symmetry broken
phase, the complex field can be decomposed into radial and transverse
modes. The radial (or massive) mode $\rho$ is effectively frozen by
its finite mass $m_{m}\propto 2\tilde{\lambda}_{k}\tilde{\kappa}_k$,
while the remaining massless Goldstone mode ($m_{g}=0$) is effectively
described by the spin-wave Hamiltonian \eqref{Eq2} and has algebraic
correlations.  On the other hand, for initial conditions in the small
$\tilde{\kappa}_{k}$ region, the flow is rapidly attracted to the
point $\tilde{\kappa}_{k}=\tilde{\lambda}_k=0$ and enters a
high-temperature $U(1)$ symmetric phase with exponential correlations,
which is identified with the disordered, high-temperature phase of the
BKT transition.

It is remarkable that the FRG treatment of the $O(2)$ model is able to
recover the high-temperature phase without explicitly considering
vortex configurations. Indeed, the complex field parametrization (i)
implicitly includes the compact phase variable responsible for vortex
excitations, in contrast to the spin-wave action \eqref{Eq2} when the
phase is considered non-periodic.

On the other hand, in the thermodynamic limit $k\to0$ there is only
one regime in Fig.~\ref{Fig1}, showing that spin-wave excitations are
always massive in this approximation. More precisely, the FRG flow
presented in Fig.~\ref{Fig1} does not exhibit a sharp BKT transition
but rather a smooth crossover.  Indeed, for large enough length scales
$k^{-1}\gg a$ the flow always reaches the symmetric phase and
algebraic correlations disappear in the thermodynamic limit for any
$T>0$. This is the result of vortex unbinding, hence the FRG
calculation \cite{Grater1995} overestimates the effect of vortex
configurations which appear to be relevant at any finite temperature.

Note that a similar behavior was already found in the Migdal approach
to the $XY$ model \cite{Jose1977}.  There, the RG equations are
written in terms of the periodic potential $V(\theta)$ between the
phases of two neighboring spins. Even in that scheme an unstable
pseudo-fixed line is found with a phase potential very similar to the
one of the Villain model \cite{Villain1975}.  On the other hand, for
small enough values of $k$ the interaction potential always reaches a
high-temperature fixed point.

The failure of the Migdal approximation to reproduce the expected
low-energy physics of the $2d$ $XY$ model has been attributed to an
insufficient representation of vortex correlations, which leads to a
systematic overestimation of the vortex contribution in the
long-wavelength limit \cite{Jose1977}.  A similar effect may be
responsible for the picture found in the lowest-order FRG
truncation. Indeed, neglecting higher derivative terms in the
$|\varphi|^4$ action may overestimate the effect of vortex degrees of
freedom in the thermodynamic limit. Nevertheless, it is an open
question whether fully including higher derivatives reproduces
vortex-vortex correlations with the correct power-law decay to
stabilize the low-temperature phase.

\begin{figure}[ht!]
\includegraphics[width=0.45\textwidth]{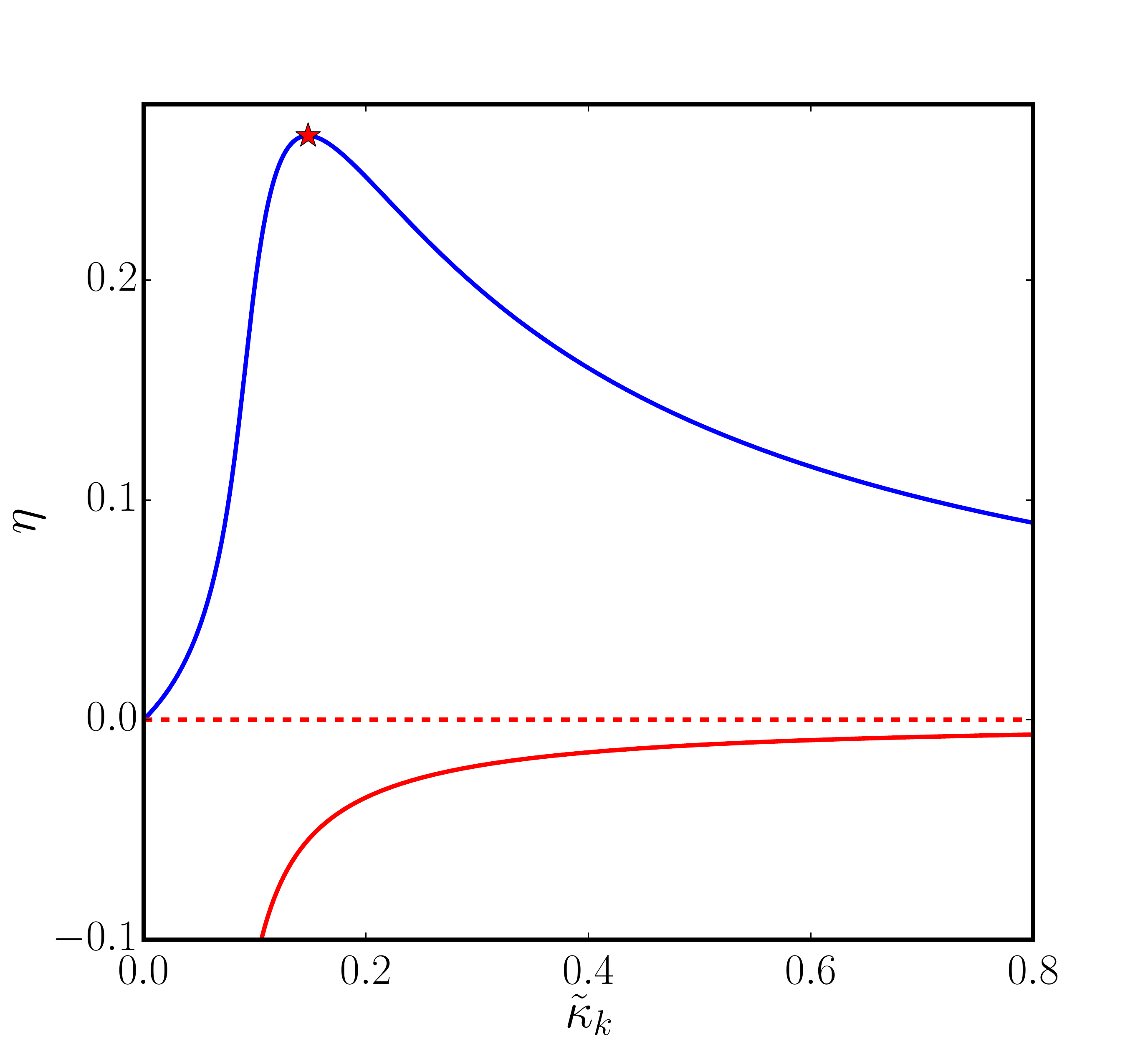}
\caption{The anomalous dimension $\eta$ in the $O(2)$ model (blue
  solid line) represents the power-law decay of the two-point
  correlation function.  Its value is found from Eq.~\eqref{Eq15}
  along the pseudo-fixed line in Fig.~\ref{Fig1}.  The star represents
  the choice of $\eta$ in \cite{Grater1995} and \cite{Jakubczyk2014}.
  The lower, red solid curve gives the values of
  $\partial_{t}\tilde{\kappa}_{k}$ along the same pseudo-fixed line,
  while a line of true fixed points would have
  $\partial_{t}\tilde{\kappa}_{k}=0$, as one has in
  \cite{Jakubczyk2014} by a temperature-dependent choice of the cutoff
  function.}
\label{Fig2}
\end{figure}

One way to extract an anomalous dimension from the LPA$'$ FRG
treatment with complex field parametrization (i) is to effectively
discard the finite flow along the pseudo-fixed line.  The condition
$\partial_{t}\tilde{\lambda}_{k}=0$ is evaluated numerically to obtain
a curve $\tilde{\lambda}_k=f(\tilde{\kappa}_{k})$ in
$(\tilde{\kappa}_{k},\tilde{\lambda}_k)$ space.  Once the finite flow
$\partial_t\tilde{\kappa}_k$ is discarded along this line, one may
compute the power-law exponent $\eta$ of the correlation function in
the thermodynamic limit using Eq.~\eqref{Eq15}.

The result for $\eta$ along the line of pseudo-fixed points is
depicted as a blue line in Fig.~\ref{Fig2}. The red curve below shows
the residual flow $\partial_t \tilde{\kappa}_k$ along the pseudo-fixed
line.  This flow should vanish for a line of true fixed points, while
as it can be seen in in Fig.~\ref{Fig2} it vanishes only in the limit
$\tilde{\kappa}_{k}\to\infty$, remaining finite for smaller
$\tilde{\kappa}_k$.  Nevertheless, it is possible to identify a point
where $|\partial_t \tilde{\kappa}_k|$ starts to increase sharply and
drives the system to the disorder phase for small scales $k$.  The
anomalous dimension at the turning point is surprisingly close to the
expected value $1/4$. In the Sections \ref{Sec4:AP}--\ref{sec:res}
below we show how a line of \emph{true} fixed points and gapped
amplitude excitations are found with the AP parametrization.  As a
basis for this, we first discuss the mapping between the $XY$ and
$|\varphi|^4$ models in Section \ref{sec:mapping}.

\section{Mapping of the models}
\label{sec:mapping}

In this section we derive the explicit mapping of the $XY$ model into
a suitable $|\varphi|^4$ theory via a Hubbard-Stratonovich
transformation \cite{Simanek1994, Machado2010, Nishimori2011}. While
this mapping is well known, we present it briefly in order to
demonstrate how the XY model of unitary spins is equivalent to a
complex field $\varphi$ with density fluctuations.  Via the mapping,
our subsequent FRG analysis of the $|\varphi|^4$ model applies also to
the XY model.

Our starting point is the $XY$ model \eqref{eq:XY}, which can be
written---apart from a constant energy---as
\begin{align}
H_{XY}=-J\sum_{\langle ij\rangle}\left(s_{x,i}s_{x,j}+s_{y,i}s_{y,j}\right),
\end{align}
where $s_{x,i}\equiv \cos\theta_i$, $s_{y,i} \equiv \sin\theta_i$ can
be combined into a vector $\vec s_i=\left(s_{x,i}, s_{y,i}\right)$
with $\vec s_i^{2}=1$.  The partition function is then given by
\begin{align}
Z(\beta)=\int \mathcal{D}s\,e^{\beta J\sum_{\langle ij\rangle}\left(s_{x,i}s_{x,j}+s_{y,i}s_{y,j}\right)}\Pi_{j}\delta\left(\vec{s}_{j}^{\,2}-1\right)
\end{align}
with $Ds=\Pi_{i}ds_{x,i}ds_{y,i}\equiv \Pi_{i} d\vec{s}_{i}$.  One can
rewrite the partition function in the form
\begin{align}
  \label{eq:ZK}
Z(\beta)=\int \mathcal{D}s \, e^{\boldsymbol{s}\cdot \frac{K'}{2} \cdot \boldsymbol{s}}\Pi_{j}\delta\left(\vec{s}_{j}^{\,2}-1\right)
\end{align}
where $\boldsymbol{s}=(s_{x,1},s_{y,1},\cdots,s_{x,N},s_{y,N})$ is a
$2N$-dimensional vector and the matrix $K'$ has elements $2 \beta J$
on the neighboring upper and lower diagonals. To perform the
Hubbard-Stratonovich transformation we use the Gaussian identity
\begin{align}
e^{\boldsymbol{s}\cdot \frac{K'}{2} \cdot \boldsymbol{s}}=\left[(2\pi)^{N}\sqrt{\det K'}\right]^{-1}\int \mathcal{D}\phi \, 
e^{-\boldsymbol{\phi}\cdot \frac{K'^{-1}}{2}\cdot \boldsymbol{\phi}-\boldsymbol{s}\cdot \boldsymbol{\phi}}
\end{align}
where $\boldsymbol{\phi}$ is a $2N$ vector composed of $N$
two-component vectors $\vec{\phi}_{j}$.  Since $K'$ is not positive
definite, we replace it by a shifted interaction
\begin{align}
\label{EqB10}
K=K'+2\beta\mu\,\mathbb{I} 
\end{align}
that is positive definite for an appropriately chosen 
constant $\mu$; this amounts to a redefinition of
the zero point energy of the system.
We then obtain 
\begin{align}
\label{EqB11}
Z(\beta)=\left[(2\pi)^{N}\sqrt{\det K}\right]^{-1}\int\mathcal{D}\phi\,
e^{-\boldsymbol{\phi}\cdot \frac{K^{-1}}{2}\cdot \boldsymbol{\phi} + \sum_{j}U(\vec{\phi}_{j})},
\end{align}
where the potential $U$ is defined by 
\begin{align}
\label{EqB12}
e^{U(\vec{\phi}_{j})}&=\int d\vec{s}_{j}\,e^{-\vec{s}_{j} \cdot \vec{\phi}_{j}} 
\delta\left(\vec{s_j}^{2}-1\right).
\end{align}
$U$ can depend only 
on the quadratic invariant $\rho_j=\phi_{x,j}^{2}+\phi_{y,j}^{2}$, and we obtain
\begin{align}
\label{EqB18}
U(\vec{\phi})=\log\left(\pi I_0\left(\sqrt{\rho}\right)\right)
\end{align}
in terms of the modified Bessel function $I_0$.
The matrix $K^{-1}$ is diagonal in Fourier space with entries
\begin{align}
\label{EqB19}
K(q)=2\beta(\mu+J\varepsilon_{0}(q))
\end{align}
where
\begin{align}
\label{EqB20}
\varepsilon_{0}(q)=\sum_{\nu=1}^{d}\cos(q_{\nu}a)
\end{align} 
is the dispersion relation on a $d$-dimensional cubic lattice for
momentum components $q_{\nu}$ and lattice spacing $a$ (in our case $d=2$). 
It will be convenient to shift the kinetic term as
\begin{align}
\label{EqB21}
S_\text{kin}[\phi]=\frac{1}{2}\sum_q\vec{\phi}_{q}\left(\frac{1}{K(q)}-\frac{1}{K(0)}\right)\vec{\phi}_{-q}.
\end{align}
After a field rescaling
\begin{align}
\label{EqB22}
\vec{\phi}\to 2\sqrt{\frac{\beta}{J}}(Jd+\mu)\vec{\varphi}
\end{align}
one obtains the kinetic term
\begin{align}
\label{EqB23}
S_\text{kin}[\varphi]=\frac{1}{2}\sum_q\vec{\varphi}_{q}\varepsilon(q)\vec{\varphi}_{-q}
\end{align}
with dispersion relation\cite{Machado2010}
\begin{align}
\label{EqB24}
\varepsilon(q)=2(Jd+\mu)\frac{d-\varepsilon_{0}(q)}{J\varepsilon_{0}(q)+\mu}.
\end{align}
In the continuum limit $a\to 0$ we recover
\begin{align}
\label{EqB25}
\varepsilon_\text{cl}(q)=q^{2}
\end{align}
to lowest order in $q$, where the subscript ``cl'' stands for
continuum limit.  The potential term in the rescaled field reads
\begin{align}
\label{EqB26}
S_\text{pot}[\varphi]=\int d^dx \left[-U\left(2\sqrt{\frac{\beta}{J}}(Jd+\mu)|\vec{\varphi}|\right)+\frac{Jd+\mu}{J}|\vec{\varphi}|^{2}\right].
\end{align}
With this mapping of the $XY$ model into a $|\varphi|^4$ theory, we
can subsequently use our functional RG equations for both models; only
the initial conditions, \emph{i.e.}, the functional forms of the
dispersion $\varepsilon(q)$ and the potential $U(\rho)$, are different
and discriminate between the microscopic $XY$ and $|\varphi|^4$
models. We finally observe that in the $XY$ model, amplitude
fluctuations are absent by construction, but there are spin-wave
excitations which interact with vortex fluctuations to modify the
  effective phase stiffness in the thermodynamic limit.  In the $O(2)$
  equivalent \eqref{EqB26} of the XY model, the finite renormalization
  of the stiffness is partly due to the gapped amplitude fluctuations,
  and partly due to longitudinal phase fluctuations, too.  Hence,
  amplitude fluctuations originate from the re-parametrization of the
  interaction between vortex and spin-wave fluctuations and, as it
  will be shown in the following, represent a large part of the
  renormalization of the superfluid stiffness.

\section{The Amplitude-Phase parametrization}
\label{Sec4:AP}

The complex field $\varphi$ in \eqref{Eq13}, which is equivalent to
  the two-component field $\vec{\varphi}$ in \eqref{EqB26}, can be
  parametrized in terms of real amplitude $\rho$ and phase $\theta$
  according to
\begin{align}
\varphi(x)=\sqrt{\rho(x)}e^{i\theta(x)}.
\label{AP}
\end{align} 
In this AP parametrization (iii) the $|\varphi|^4$ action
\eqref{Eq13} reads
\begin{align}
\label{eq:SAP}
S[\varphi]=\int{d^{d}x\left\{\frac{1}{8\rho}\partial_{\mu}\rho\partial_{\mu}\rho+\frac{\rho}{2}\partial_{\mu}\theta\partial_{\mu}\theta+U(\rho)\right\}}.
\end{align}
When applied to the $XY$ model with the mapping \eqref{EqB26}, the
field expectation value is related to the $XY$ magnetization by
$\langle\varphi\rangle = \sqrt{\beta J} m$.

It is worth noting that the derivation of action \eqref{eq:SAP} is a
direct consequence of Leibniz rule for continuous spatial derivatives
and it is not straightforwardly applicable to lattice models where the
kinetic contribution to the action cannot be expressed in terms of
continuous spatial derivatives. In any case, lattice effects can be
introduced by integrating the model in the $\varphi$ parametrization
till an effective scale where only quadratic momentum terms dominate,
as it should always happen due to universality.

Perturbative arguments suggest that the amplitude mode is always
gapped and does not influence the critical
behavior\cite{Popov1983,Jakubczyk2017}.  Instead, the critical
behavior is dominated by massless phase fluctuations.  Indeed, in
$d=2$ only single vertex diagrams are relevant \cite{Rajaraman1987},
and since the perturbative expansion for the phase correlation
function does not contain any single vertex diagram, we expect only a
finite renormalization of the superfluid stiffness in the action
\eqref{eq:SAP}.  In the following we will explicitly treat amplitude
fluctuation effects to show how, even in the nonperturbative picture,
they remain gapped at criticality.

Consistent with these arguments, we propose a RG investigation which
treats amplitude and phase fluctuations separately in two steps.  The
overall RG flow is thus effectively separated into two scales: first,
at high momenta, only non-critical fluctuations are considered.  They
turn out to be irrelevant and remain always gapped in the
long-wavelength limit.  At the end of the AP flow the minimum
$\kappa_k$ freezes, and we obtain an effective SG model with a
renormalized superfluid stiffness.  The effective couplings resulting
from the amplitude stage of the flow are then considered as initial
conditions for the traditional BKT flow\,\cite{Kosterlitz1973} of the
SG model, which describe low-energy vortex excitations and produce the
universal long-wavelength behavior.  The validity of our technique
relies on the separation of the scales for amplitude and vortex phase
fluctuations, and on the smallness of the interaction between these
excitations.  Traditional perturbative arguments as well as the
consistency of our results suggest that these assumptions are well
justified.

As a preliminary step, we first discuss uncoupled amplitude and phase
fluctuations.  In this case, the superfluid stiffness
$\rho=\kappa_{k}$ in the phase kinetic term remains fixed at the
minimum $\rho_0$ of the potential $U(\rho)$.  The total action \eqref{eq:SAP}
then decouples into a sum of two actions
\begin{align}
\label{Eq18}
S[\varphi]\simeq S_A[\rho]+S_P[\theta]
\end{align}
where
\begin{align}
S_A[\rho]&= \int{d^{2}x\,\left\{\frac{1}{8\rho}\partial_{\mu}\rho\partial_{\mu}\rho+U(\rho)\right\}},\label{eq:Srho}\\
S_P[\theta]&= \frac{\kappa_{k}}{2} \int{d^{2}x \, \partial_{\mu}\theta\partial_{\mu}\theta}.\label{eq:Sphase}
\end{align}
The phase action \eqref{eq:Sphase} is equivalent to spin-wave model
\eqref{Eq2} with $K=\kappa_{k}$, while for the $XY$ model it is
$K=\kappa_{k}\beta J$.  If one considers the phase variable $\theta$
in \eqref{eq:Sphase} as noncompact, the correlation function
$\langle e^{i[\theta(x)-\theta(y)]}\rangle$ is
algebraic\cite{Berezinskii1972} and no regularization is necessary to
obtain this behavior\cite{Jakubczyk2017}.

The treatment within the AP parametrization shows that the
low-temperature expansion of the $|\varphi|^4$ and $XY$ models must
coincide, at least as long as perturbative arguments are correct and
amplitude fluctuations do not influence the thermodynamic behavior.
However, it is worth noting that this analysis still does not yield a
conclusive picture.  Indeed, while the previous FRG analysis based on
the $|\varphi|^4$ action \eqref{Eq13} leads to a finite correlation
length at any temperature and reproduces the BKT behavior only as a
crossover, the amplitude and phase scheme is equivalent to the
spin-wave approximation of the $XY$ model and yields algebraic
correlation at any temperature, $T_\text{BKT}=\infty$.

To complete the picture, it is therefore necessary to introduce vortex
configurations.  The spin-wave analysis in Appendix~\ref{AppendixA}
does not include discontinuous configurations of the field $\theta$
and perturbative arguments cannot account for topological excitations.
These can be included using the dual mapping described in
\cite{Jose1977,Kleinert1989} or by explicitly introducing singular
phase configurations \cite{Mudry2014,Benfatto2012}.  The total
partition function of the system is then given by
\begin{align}
\label{Eq22}
Z\simeq Z_{A}Z_{P},
\end{align}
where we used the decomposition in Eq.~\eqref{Eq18}.

In the case of frozen amplitude fluctuations, this model becomes a
pure phase SG model with a line of fixed points and is described by
the BKT flow equations
\begin{align}
\partial_{t}K_{k}&=-\pi g_k^{2}K_k^{2},\label{Eq23}\\
\partial_{t}g_{k}&=\pi\left(\frac{2}{\pi}-K_k\right)g_k\label{Eq24}
\end{align}
where $K$ is the superfluid (phase) stiffness and $g_{k}$ is the
vortex fugacity.  The fugacity $g$ is related to the SG parameter as
$u=g/\pi$.

At the bare level, $K$ and $g$ assume the values
\begin{align}
K_{\Lambda}&=\rho_0,\label{Eq25_0}\\
g_{\Lambda}&=2\pi e^{-\pi^{2}K_{\Lambda}/2}\label{Eq26_0}
\end{align}
for the $|\varphi|^4$ model, and 
\begin{align}
K_{\Lambda}&=\beta J,\label{Eq25}\\
g_{\Lambda}&=2\pi e^{-\pi^{2}K_{\Lambda}/2}\label{Eq26}
\end{align}
for the $XY$ model. In order to derive Eqs.~\eqref{Eq23}--\eqref{Eq24}
one has to assume a UV regularization, which traditionally relies in
considering the Coulomb gas charges as hard disks of finite
radius\cite{Kosterlitz1973}.

It should be also noted that Eqs.~\eqref{Eq25_0}--\eqref{Eq26}
have been obtained in the case of a purely quadratic kinetic phase
term, as in the Villain model.  In the $O(2)$ model the absence of
higher gradient terms in the phase is the result of the decoupling
in equation \eqref{Eq18}.  In principle, one expects amplitude
fluctuations to generate also higher gradient terms and therefore
action \eqref{eq:Sphase} represent only the lowest-order
approximation in derivative expansion.

In the small vortex fugacity limit $g_{k}\ll 1$ the BKT flow
Eqs. \eqref{Eq23} and \eqref{Eq24} reproduce the BKT temperature in
Eq.~\eqref{Eq7}, while for larger values of the initial condition
$g_{\Lambda}$ the BKT flow introduces multi-vortex corrections which
lower the BKT temperature.  For a discussion of these effects and of
vortex core energies we refer to \cite{Prokofev2000,Hasenbusch2005};
the prediction for the jump of the superfluid stiffness,
$\frac{2mT}{\pi} \left( 1-16\pi e^{-4\pi}\right)$ with a correction of
$0.02\%$ with respect to the Nelson-Kosterlitz prediction $2mT/\pi$,
has been tested in extensive Monte Carlo simulations
\cite{Hasenbusch2005,Komura2012}.

\section{Results} 
\label{sec:res}

Although the universal behavior of the BKT transition is completely
driven by topological excitations, in the $|\varphi|^4$ and $XY$
models the contribution of, respectively, longitudinal and amplitude
fluctuations to non-universal quantities may be different.  Due to the
mapping discussed in Section \ref{sec:mapping}, it is possible to
build a $|\varphi|^4$ model which exactly reproduces the $XY$ model
and where amplitude fluctuations play the role of longitudinal spin
excitations.  It is then convenient to study the BKT transition first
in the $|\varphi|^4$ formalism and then transfer the results to the
$XY$ model, which we do subsequently in the two next Sections
\ref{sec:phi4_r} and \ref{sec:XY_r}.

\subsection{$\bm{|\varphi|^4}$ model}
\label{sec:phi4_r}

In this section we apply the FRG to the $|\varphi|^4$ action in the AP
parametrization \eqref{eq:SAP}.  As discussed in the previous section,
at the perturbative level the amplitude mode $\rho$ remains gapped
while the phase fluctuations $\theta$ produce power-law correlations
at any finite temperature, so the high-temperature phase of the BKT
transition is not reproduced.  In this section we revisit this issue
at the nonpertubative level.

Our FRG procedure is based on two steps: (a) we first perform the FRG
flow for the amplitude part $S_A$ of the action \eqref{eq:Srho}, which
yields a renormalized superfluid stiffness; (b) we then insert this
stiffness into the phase part $S_P$ of the action \eqref{eq:Sphase},
which for a compact phase is equivalent to the SG model \eqref{eq:SG}
so we can use the BKT flow Eqs.~\eqref{Eq23}--\eqref{Eq24}.
\begin{figure*}[t!]
\centering
\subfigure[\large]{\label{Fig3a}\includegraphics[width=.45\textwidth]{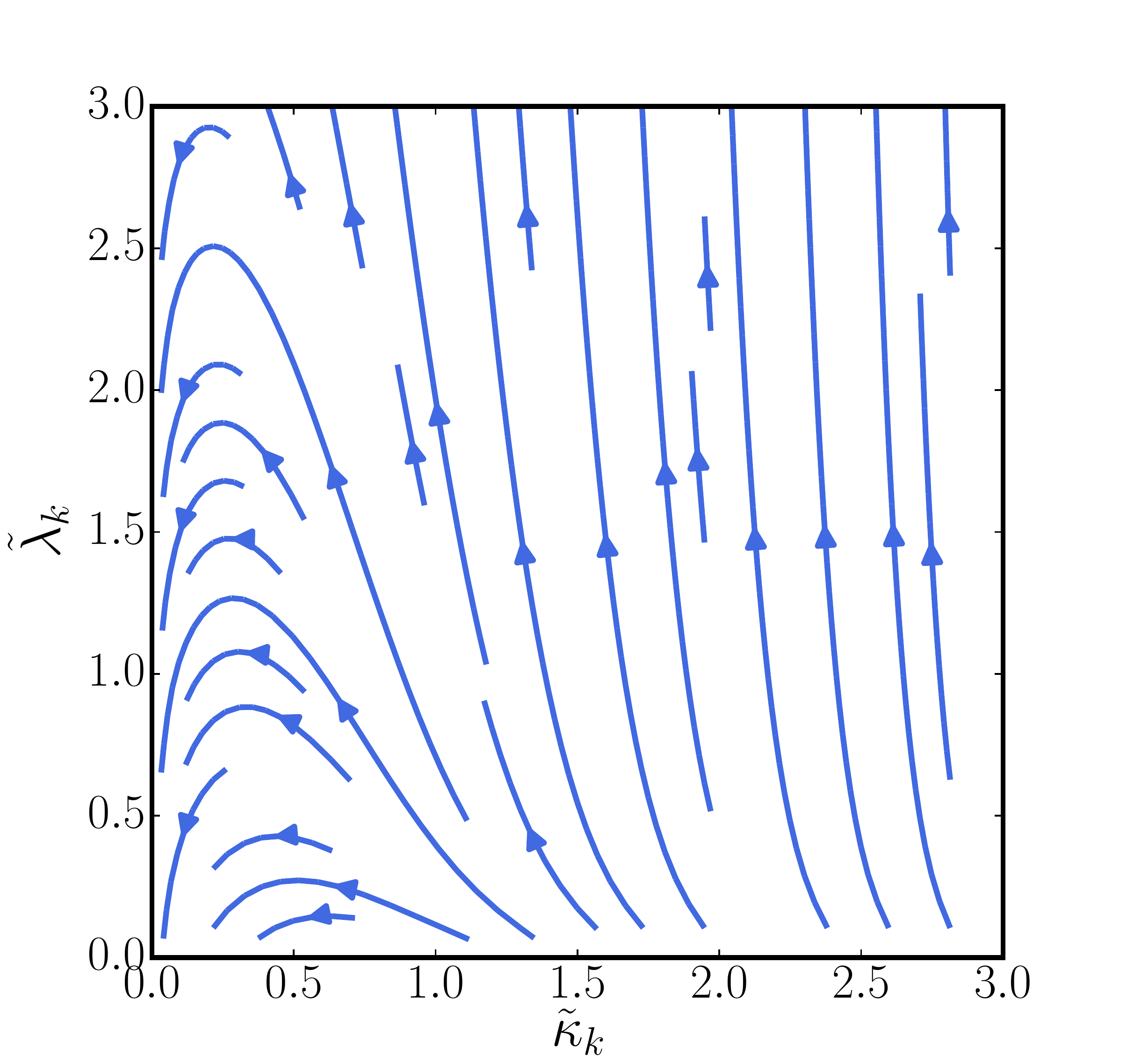}}
\subfigure[\large]{\label{Fig3b}\includegraphics[width=.45\textwidth]{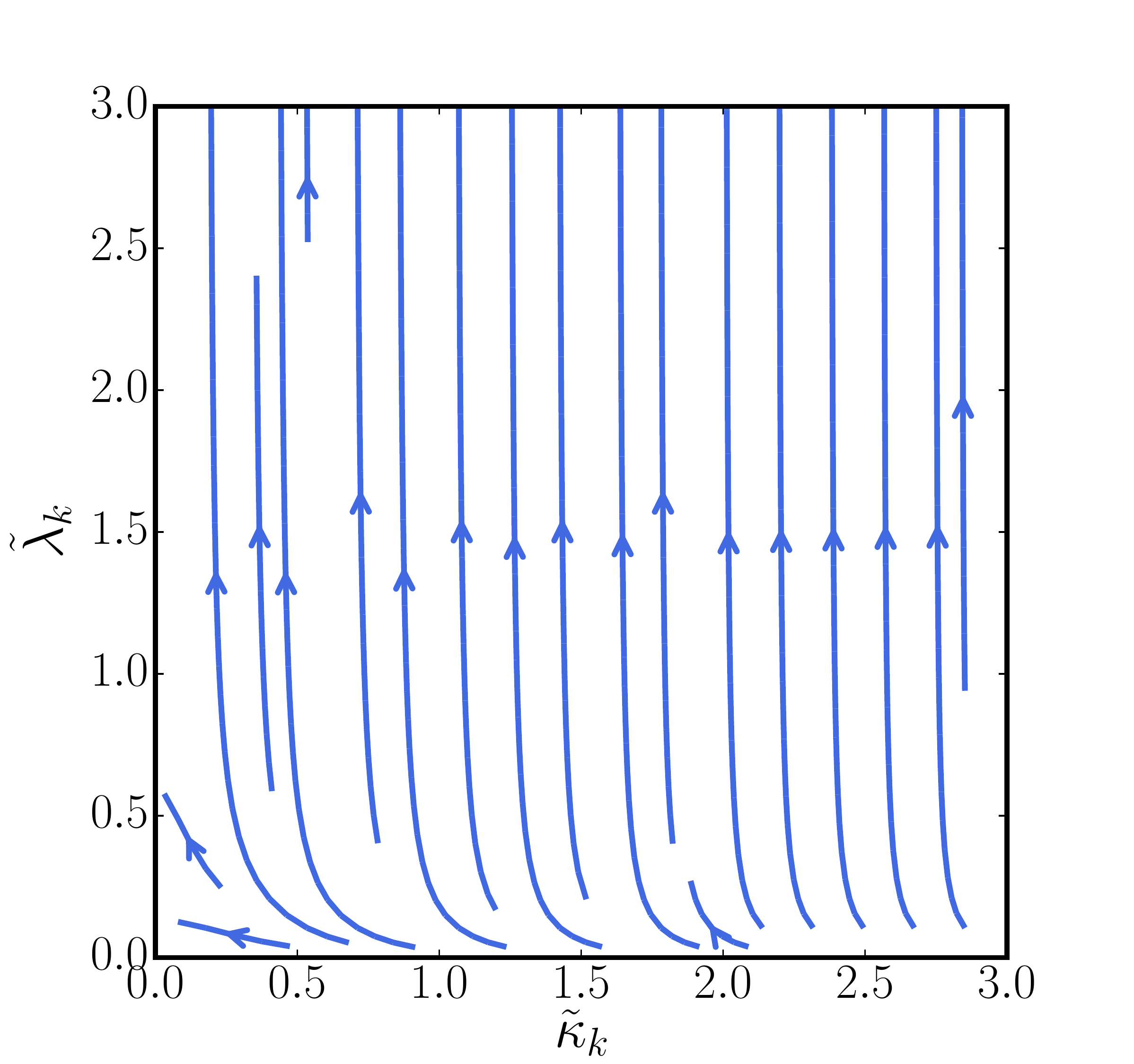}}
\caption{Flow diagram for the rescaled superfluid stiffness 
$\tilde{\kappa}_{k}$ and interaction $\tilde{\lambda}_{k}$ due to 
amplitude fluctuations in $d=2$. For large enough $\tilde{\lambda}_{\Lambda}$ 
the flow always proceeds towards an infinitely interacting 
$\tilde{\lambda}_{k\simeq0}\simeq+\infty$ fixed line where
the expectation value $\tilde{\rho}_{0}=\kappa_{k}$ is effectively frozen. 
\textbf{(a)} The naive AP flow is (uncorrectly) 
attracted for $\tilde{\lambda}_{\Lambda}\ll 1$ toward the free theory. 
\textbf{(b)}
The modified AP flow with the Gaussian 
contributions subtracted reproduces the expected flow diagram.}
\label{Fig3}
\end{figure*}

In the FRG approach for the amplitude part we introduce as infrared
regulator a momentum dependent mass term for the amplitude
fluctuations.  As the cutoff scale is lowered, the effective action
flows from the model-dependent initial condition \eqref{eq:Srho} to
the full effective action.  For the flowing effective action we choose
the ansatz
\begin{align}
\label{eq:Gammak}
\Gamma_{k}[\rho,\theta]=\int{d^{d}x\left\{\frac{1}{8\rho}\partial_{\mu}\rho\partial_{\mu}\rho+\frac\rho2\partial_{\mu}\theta\partial_{\mu}\theta+U_{k}(\rho)\right\}},
\end{align}
and with the regulator \eqref{EqC2} we obtain the flow
Eq.~\eqref{EqB6} for the effective potential of amplitude and phase
fluctuations, for details see Appendix~\ref{AppendixC}.

The flow equation is solved numerically for the full potential
$U_k(\rho)$.  In order to draw a flow diagram, we Taylor expand the
potential $U_k=\lambda_k (\rho-\kappa_k)^2/2$ around its minimum
$\rho=\kappa_k$ for every $k$ and trace the flow in
$(\kappa_k,\lambda_k)$ space.  The resulting flow diagram is shown in
the left Fig.~\ref{Fig3a} in terms of the rescaled ``dimensionless''
couplings $\tilde{\lambda}_{k}$ and $\tilde{\kappa}_k$.  This first
naive attempt at the AP flow is not yet correct: indeed in the lower
left corner of the phase diagram the $\lambda_{k}$ coupling becomes
irrelevant and the flow runs toward a region of gapless amplitude
fluctuations; although this effect is not as severe as in previous
parametrization, since it arises only for small values of the bare
coupling $\lambda_{\Lambda}$, it is not in agreement with the
expectation of irrelevant amplitude fluctuations in the thermodynamic
limit.  This inconsistency arises from an IR divergent term in the
standard formulation of the Wetterich equation. Indeed, already the
flow of the free Gaussian model in the AP parametrization has the same
divergence because the phase kinetic term depends on the field $\rho$.

This spurious contribution originates from the different ways the
  Gaussian theory is represented in the amplitude and phase
  parametrization.  In the path integral formulation the Gaussian
  $O(2)$ model, \eqref{Eq13} with $U=0$, can be exactly integrated,
  yielding an effective action with the same functional form of the
  microscopic action, in agreement with mean field approximation being
  exact for Gaussian theories.  Thus also in the FRG formalism the
  flow of Gaussian theories should vanish and the bare action should
  be equivalent to the fully renormalized one.  Nevertheless when one
  uses the Wetterich equation to calculate the flow of a Gaussian
  ansatz in the traditional $O(N)$ formalism one gets a constant flow
  for the effective action \cite{Krippa2009, Gies2012}.  Thus, the
  functional form of a Gaussian action is preserved by the FRG
  formalism apart for a field independent term, which is diverging in
  the long-wavelength limit.  Such a term is unnecessary in the
  computation of most of the system properties and it is usually
  neglected.

In free energy calculations the constant term in the effective
  action is essential and should be kept finite.  The most common
  regularization procedure is to subtract the noninteracting
  ($\lambda_k=0$) component from the right-hand side of the Wetterich
  equation under study \cite{Krippa2009, Gies2012}.  The Gaussian
  theory in the AP parametrization has a linear field potential term
  and a nonanalytic kinetic term, as in Eq.~\eqref{eq:Gammak} with
  $U_{k}(\rho)=\mu\rho$.  The nonanalytic term prevents the exact
  integration in the AP representation and makes it appear to be not
  exactly solvable; in fact, the AP representation is singular in the
  $\rho\to0$ limit.  In our application the amplitude fluctuations
  remain gapped, so the AP parametrization is always valid.

As stated above, the FRG flow of Gaussian theories does not
  completely vanish, and in the amplitude and phase representation the
  remaining contribution also pick up a spurious field
  dependence, which is the counterpart of the nonanalytic kinetic term
  preventing an exact integration in the path integral formalism.
  Since one knows that the functional form of quadratic theories
  remains the same at bare and renormalized level and this property
  must remain valid regardless of the chosen representation, one can
  safely subtract the equivalent Gaussian contribution from the flow
  equations and force them to be zero for quadratic theories, as it is
  done in the traditional case for free energy calculations.

With this modification the potential flow equation 
\eqref{EqB6} becomes
\begin{align}
\partial_{t}U_{k}(\rho)=\frac{4\alpha  \rho k^2 \log \left(\frac{\alpha +4 \alpha  \rho  U^{(2)}(\rho)/k^2}{\alpha + U^{(2)}(\rho)/k^2}\right)}{4 \pi  (4\alpha  \rho -1)}.\label{eq:Uflow}
\end{align}
The parameter $\alpha=\alpha_{\rho}$ characterizes the regulator
\eqref{EqC2}; Fig.~\ref{Fig3} has been plotted with $\alpha_{\rho}=2$,
but below we set $\alpha=(4\kappa_{*})^{-1}$ self-consistently with
the value of $\kappa$ at the end of the flow.

The modified Eq.~\eqref{eq:Uflow} now produces the correct flow
diagram shown in the right Fig.~\ref{Fig3b}.  The flow
  equations for the dimensionless minimum $\tilde\kappa_{k}$ and
  interaction $\tilde\lambda_{k}$ read
\begin{align}
\partial_{t}\tilde{\kappa}_{k}=&\,\frac{\alpha\left(\frac{4 \tilde{\kappa}_{k} (1-4 \alpha  \tilde{\kappa}_{k})}{4 \tilde{\kappa}_{k}\tilde{\lambda}_{k}+1}+\frac{\log \left(\frac{\alpha +4 \alpha  \tilde{\kappa}_{k} \tilde{\lambda}_{k}}{\alpha +\tilde{\lambda}_{k}}\right)}{\tilde{\lambda}_{k}}\right)}{\pi  (1-4 \alpha  \tilde{\kappa} _{k})^2}\\
\partial_{t}\tilde{\lambda}_{k}=&\,2\tilde{\lambda}_{k}-\frac{8 \alpha ^2 \log \left(\frac{\alpha +\tilde{\lambda}_{k}}{\alpha +4 \alpha  \tilde{\kappa}_{k}\tilde{\lambda}_{k}}\right)}{\pi  (4 \alpha\tilde{\kappa}_{k}-1)^3}\nonumber\\
&-\frac{8 \alpha 
   \tilde{\lambda}_{k} (2\tilde{\kappa}_{k} \tilde{\lambda}_{k} (4 \alpha\tilde{\kappa}_{k}+1)+1)}{\pi  (1-4 \alpha \tilde{\kappa}_{k})^2 (4 \tilde{\kappa}_{k} \tilde{\lambda}_{k}+1)^2}.
\end{align}
As expected, the mass term of amplitude fluctuations does not vanish.
The dimensionless $\tilde\lambda_k$ keeps growing because the action
\eqref{eq:Gammak} with a noncompact phase has no fixed point. Indeed,
$\tilde\kappa_{k}$ is marginal in $2d$, and after an initial
renormalization by amplitude fluctuations at finite
$\tilde\lambda_{k}$, it remains frozen up to infinite length scales
($k\to0$).

The results of Fig.~\ref{Fig3b} are in agreement with the expectation
from perturbation theory and show that amplitude fluctuations are
irrelevant in the RG sense and only lead to a finite renormalization
of the stiffness.  Note that this irrelevance has been proven in
  the truncation scheme described by the ansatz \eqref{eq:Gammak},
  where only the lowest-order coupling between amplitude and phase is
  present.  Indeed, also phase fluctuations drive the flow of the
  effective potential $U_k(\rho)$ in the first term of the original
  flow equation \eqref{EqB6}, but this contribution is canceled when
  subtracting the Gaussian part to obtain \eqref{eq:Uflow}.  In a more
  general truncation scheme we do not expect this cancellation to
  persist, but we are confident that the remaining phase-amplitude
  terms will be irrelevant.

It is useful to compare the results of Figs.~\ref{Fig1} and
\ref{Fig3a} with those of Fig.~\ref{Fig3b}: both represent the theory
space of a $2d$ two-component field theory where the order parameter
has $U(1)$ symmetry. However, differences arise in the treatment of
the kinetic term: in Fig.~\ref{Fig1} the flow for the couplings has
been obtained including the full $|\varphi|^4$ invariant kinetic term,
which incorporates both amplitude and phase degrees of freedom.
There, for $\lambda_{\Lambda}$ large enough, the flow is attracted to
a pseudo-fixed line and the IR theory appears to have finite
$\tilde{\kappa}_{k}$, finite $\tilde{\lambda}_{k}$ and massive
amplitude fluctuations at finite $k$.  At the same time, a fixed
$\tilde{\lambda}_{k}$ produces a vanishing dimensionful
$\lambda_{k}=k^{2}\tilde{\lambda}_{k}$ in the thermodynamic limit
$k\to0$.  Hence, it is not surprising that for $k$ small enough the
superfluid density $\kappa_{k}$ tends to vanish because of the
increasing relevance of amplitude fluctuations.  In contrast, the
modified flow in the right Fig.~\ref{Fig3b} is consistent with fully
gapped amplitude fluctuation and frozen amplitude (superfluid
stiffness) $\kappa_{k}\equiv\kappa_{*}$. Indeed, for every finite
$\tilde{\lambda}_{\Lambda}>0$ the flow is attracted by a stream line
at fixed $\tilde{\kappa}_{k}$ and
$\tilde{\lambda}_{k}\propto k^{-2}\lambda_{*}$, yielding
$\lambda_{k}\simeq\lambda_{*}$ for $k\ll\Lambda$.

\begin{figure*}[ht!]
\centering
\subfigure[\large]{\label{Fig4a}\includegraphics[width=.45\textwidth]{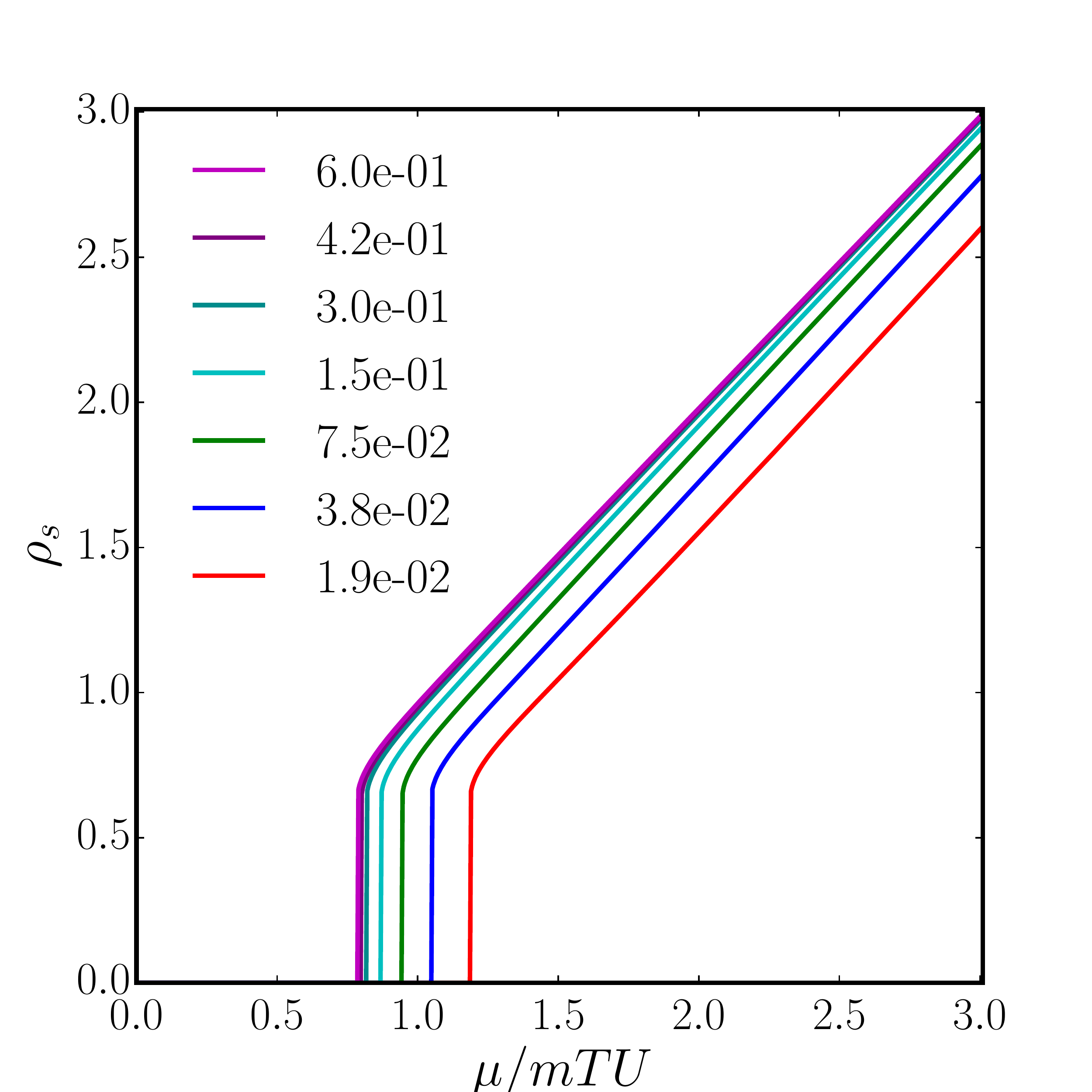}}
\subfigure[\large]{\label{Fig4b}\includegraphics[width=.45\textwidth]{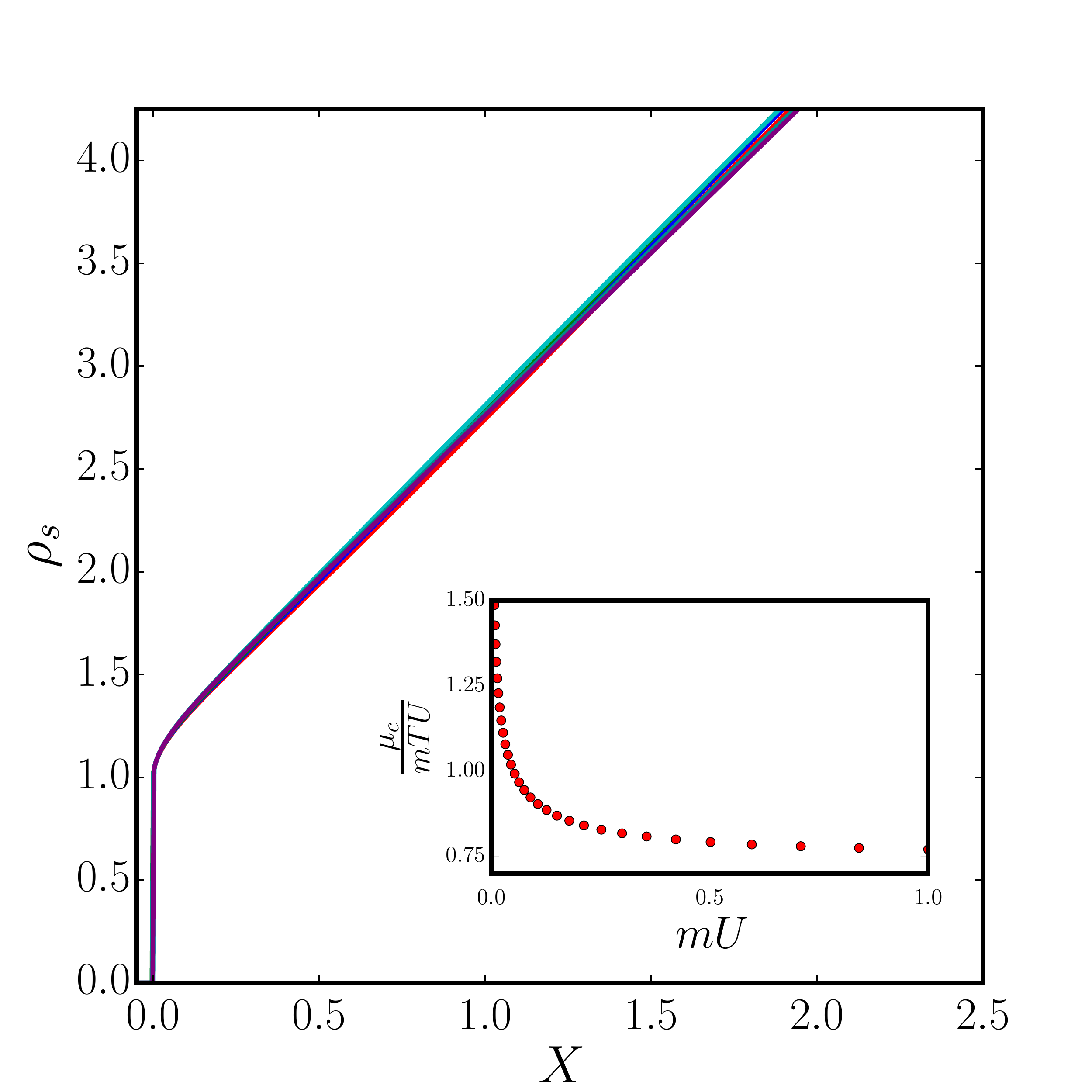}}
\caption{Superfluid density $\rho_s$ as a function of chemical
  potential.  \textbf{(a)} $\rho_s/mT$ \textit{vs} $\mu/U$ for $7$
  different values of $mU=0.6\dotsc0.02$ from top to bottom.
  \textbf{(b)} $\rho_s/mT$ \textit{vs} dimensionless chemical
  potential $X$.  \textbf{Inset:} Critical chemical potential
  $\mu_c/U$ \textit{vs} $U$.}
\label{Fig4}
\end{figure*}

Having shown that the modified AP flow agrees with the perturbative
results and the BKT scenario, we are in a position to verify that our
approach reproduces the expected universality of the thermodynamics of
the $2d$ Bose gas\cite{Popov1983, Fisher1988, Baym1999, Prokofev2001,
  Prokofev2002, Rancon2012} and to quantify the agreement with
Monte-Carlo results.  In particular, starting the flow from the
initial conditions \eqref{Eq25_0} we can compute:
\begin{enumerate}[label=\alph*.]
\item the superfluid density 
$\rho_s$, which is equal to the coupling $\kappa_{*}$; and 
\item the critical chemical 
potential $\mu_c$ as a function of the bare interaction $U$. 
\end{enumerate}
To achieve this result we perform the renormalization group procedure
described above with the initial condition
\begin{align}
U_{\Lambda}(\rho)=\frac{U}{2}(\rho-\kappa_{\Lambda})^{2},
\end{align}
where $U$ is the effective interaction and 
$\mu=U\kappa_{\Lambda}$ is the chemical potential 
of the classical $2d$ $|\varphi|^4$ model we are studying. 

The known results for the $2d$ quantum Bose gas 
with which we want to compare are the following \cite{Prokofev2001}:
\begin{itemize}
\item[1)] the thermodynamic quantities have to collapse once expressed
  in terms of the dimensionless variable
\begin{align}
X = \frac{\mu-\mu_c}{mTU},
\label{XX}
\end{align}
which measures the distance from the critical point. 

\item[2)] The superfluid density defines a function $f(X)$ via the 
relation
\begin{align}
\rho_s = \frac{2 m T}{\pi} \, f(X).
\label{effe}
\end{align}
Note that the predicted jump of the superfluid stiffness $\rho_s=2 m T / \pi$
at criticality \cite{Nelson1977} implies that $f(X)$ jumps from 0 to 1 at $X=0$.
The collapse of the superfluidity function using the variable $X$ 
is shown in Figs.~\ref{Fig4a}--\ref{Fig4b}.

\item[3)] for small $X>0$ one has 
\begin{align}
f(X) = 1 + \sqrt{2 \kappa' X},
\label{kappa_primo}
\end{align}
with coefficient\cite{Prokofev2002} 
\begin{align}
\kappa'=0.61\pm0.01.
\label{kappa_MC}
\end{align}

\item[4)] For $2d$ quantum systems in the continuum, one has the following 
results in the weakly interacting 
limit for the critical density $\rho_c$ and the critical chemical potential 
$\mu_c$ (respectively in the canonical and grand-canonical ensembles):
\begin{align}
\label{Eq47}
n_c = \frac{m T}{2 \pi} \, \ln{\frac{\xi}{mU}},
\end{align}
\begin{align}
\mu_c = \frac{m T U}{\pi} \, \ln{\frac{\xi_\mu}{mU}}.
\label{xi_mu}
\end{align}
The parameters $\xi$, $\xi_\mu$, extracted from Monte Carlo
simulations in a classical lattice $|\varphi|^4$ model and via a
careful analysis of the mapping between the simulated lattice model
and the continuum limits, have been estimated to be $\xi=380\pm 3$ and
$\xi_\mu=13.2\pm 0.4$ \cite{Prokofev2001, Prokofev2002}.  An earlier
  FRG approach, once the transition point has been empirically fixed,
  yields $\xi_{\mu}=9.48$ in good agreement with MC
  simulations\cite{Rancon2012}.  The logarithm of their ratio,
\begin{align}
\theta_0 \equiv \frac{1}{\pi} 
\ln{\left(\xi/\xi_\mu\right)},
\end{align} 
is a non-trivial universal number, determined to be 
\cite{Prokofev2002}
\begin{align}
\theta_0 = 1.068 \pm 0.01.
\label{theta0_MC}
\end{align}
\end{itemize}

\begin{figure}[t!]
\centering
\includegraphics[width=.45\textwidth]{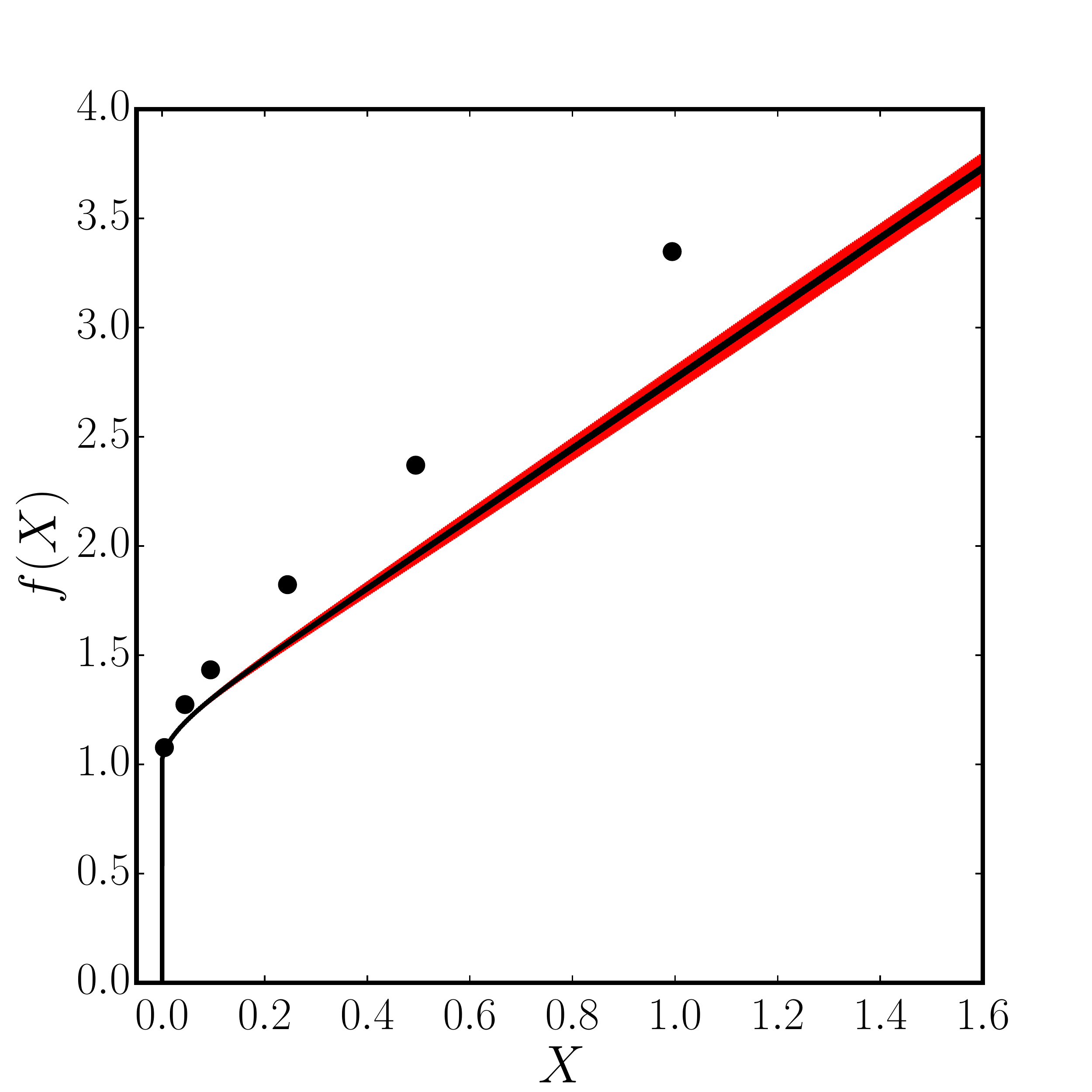}\caption{(a) Superfluid
  scaling function $f(X) = \pi \rho_s/2mT$ (black line) as a function
  of the chemical potential variable $X$ (average and variance over
  $30$ sets of data for different interaction values $U$), the
  standard deviation is shown as a red shadow. Black dots are the MC
  data from \cite{Prokofev2002}.}
\label{Fig5}
\end{figure}

We now present our results for these non-universal and universal 
properties of the $|\varphi|^4$ model. 
In Fig.~\ref{Fig4a} we report our 
results for the superfluid fraction $\rho_s$ for different values 
of $U$. In Fig.~\ref{Fig4b} we plot the same curves 
vs the dimensionless variable $X$.  
We find that they collapse almost perfectly even for a wide range of 
interactions $mU=0.02,\dotsc,0.6$. Note that the spreading between the 
curves increases for large $X$, as expected, since the universality 
should hold only in the fluctuation regime up to $X\approx 1/mU$ 
and we use also rather large values of $U$.
To quantitatively determine the function $f(X)$ 
we perform an interpolation of the curves for $\rho_{s}(X)$, 
some of them shown in Fig.~\ref{Fig4b}, and compute their average 
and variance, which are 
reported in Fig.~\ref{Fig5}. The average has been 
computed over a total number
of $30$ curves obtained for $30$ different values 
of the interaction logarithmically spaced in the interval $mU\in[0,1]$, 
the curve $f(X)$ can be trusted also for large $X$ since the statistical 
weight of large interaction $U>0.5$ is small. 
Agreement with Monte Carlo data\cite{Prokofev2002} 
is rather good, also considering that we are using the 
lowest order perturbative SG results \eqref{Eq25_0}--\eqref{Eq26_0}. 

Our findings for $\mu_c$ as a function of $U$ are given in the inset
of Fig.~\ref{Fig4b}. Logarithmic corrections to the relation
$\mu_c \propto U$ are found, in agreement with Eq.~\eqref{xi_mu}.  The
coefficient $\xi_{\mu}$ entering such logarithmic corrections is not
reported since the fitting procedure employed was not robust enough
and the result strongly depends on the range of interactions
considered, even for $U\leq 0.3$ which should be within the range of
validity of Eq.~\eqref{xi_mu} \cite{Hadzibabic2011}.

The $\rho_s(X)$ in Fig.~\ref{Fig4b} determines the function
$f(X)=\pi \rho_s(X)/2mT$ reported in Fig.~\ref{Fig5}. From $f(X)$ we
can obtain estimates for the universal quantities $\kappa'$ and
$\theta_0$. Fitting with expression Eq.~\eqref{kappa_primo}, the data
in Fig.~\ref{Fig5} yield
\begin{align}
\label{Eq50}
\kappa_\text{(FRG)}'=0.67\pm 0.07,
\end{align}
in reasonably good agreement with the Monte Carlo result
\eqref{kappa_MC}. The latter result has been obtained from a linear
fit of the curves in Fig.~\ref{Fig4a} and averaging $\kappa'$ over the
values obtained for different interactions. The average is consistent
with \eqref{Eq50} while the error is partly due to difficulties in
fitting procedure close to the transition point and partially to
non-perfect universality of the curves in Fig.~\ref{Fig4a}.

Regarding $\theta_0$, we observe that for relatively large $X$ one has
$f(X)\approx(\pi/2) \theta(X)-1/4$ in terms of the universal equation
of state $\theta(X)$ \cite{Prokofev2002}.  It should be noted that in
order to evaluate $\theta_{0}=\theta(X=0)$ from $f(X)$ one shall
extrapolate the value of a curve obtained for large $X$ to the point
$X=0$. Such extrapolation has been done assuming polynomial behavior
of $\theta(X)$. A polynomial fit of the $\rho_{s}$ curves of different
interactions at high values of $X$ yields
\begin{align}
\theta_{0\text{(FRG)}}=1.033\pm 0.032,
\end{align}
again in fairly good agreement with the Monte Carlo result \eqref{theta0_MC}. 
\subsection{$\bm{XY}$ model}
\label{sec:XY_r}

As we discussed in Sections \ref{sec:mapping}--\ref{Sec4:AP}, one can treat 
the $XY$ model as a $|\varphi|^4$ model, provided that one uses 
the appropriate initial condition for the RG flow, as extracted from the 
mapping of Section \ref{sec:mapping}, and that one rescales the field 
by $\sqrt{\beta J}$ to have a magnetization with absolute value smaller than one.

The $XY$ model has been the subject of intense investigations from
different perspectives and several quantities have been studied in
detail, which we can now study with the FRG approach presented in this
paper.  Here, to test the validity of our approach, we focus on the
renormalized phase (superfluid) stiffness $J_s(T)$ and quantify the
effect of amplitude fluctuations on it. We proceed by computing
$\kappa_{*}$ as discussed in the previous Section \ref{sec:phi4_r},
then the stiffness is given by
\begin{align}
J_{s}(T)=J\kappa_{*},
\label{kappa_star}
\end{align}
where $J_{s}(T)$ indicates the effective \emph{bare} superfluid
stiffness without inclusion of vortex configurations.  In the
following, the same notation will be used also to indicate the fully
renormalized spin stiffness in presence of vortex excitations, since
the two definitions can be simply regarded as two levels of
approximation for the same quantity.

All the physical quantities should be independent of the mapping
parameter $\mu$ \cite{Machado2010}. However, in the following we are
going to discard lattice effects, effectively replacing the lattice
dispersion \eqref{EqB24} with the continuum dispersion \eqref{EqB26}.
Such an approximation introduces a $\mu$ dependence in the physical
quantities, which we may fix either 
from mean-field or low-temperature results.

To clarify the different approximations which we are going to consider
for the FRG computation of $J_s(T)$, let us recapitulate the logic
followed so far.  Starting from the action of the $|\varphi|^4$ model
in the continuum limit, we introduced the AP parametrization
\eqref{AP} and we decoupled the phase and amplitude degrees of freedom
by substituting $\rho=\kappa_{k}$ into the phase kinetic term.  The
phase action \eqref{eq:Sphase} is then equivalent to the
low-temperature expression of the $XY$ Hamiltonian \eqref{Eq2} and we
can apply the usual BKT flow Eqs. \eqref{Eq23} and \eqref{Eq24}.  The
amplitude fluctuations then encode all fluctuations except for
vortices, which are encoded at perturbative level in the BKT flow
equations.

It is instructive to consider first the mean-field approximation.  A
first step is to completely discard amplitude fluctuation and simply
set $\kappa_{k}=\text{const}$, which can be re-absorbed into the
definition of $J$.  A further step is to consider only a saddle point
approximation for the amplitude fluctuations.  Their expectation value
is given by $\kappa_{MF}=\rho_{\mathrm{MF}}(T)$ such that
\begin{align}
\frac{\partial S_{\mathrm{pot}}[\sqrt{\beta J \rho_{\mathrm{MF}}}]}{\partial\rho_{\mathrm{MF}}}=0,
\end{align}
where $S_{\mathrm{pot}}[\varphi]$ is defined in Eq.~\eqref{EqB26} and
the additional $\sqrt{\beta J}$ factor in the argument is needed to
reproduce the $K\equiv \beta J$ factor in Eq.~\eqref{Eq2}. Thus, at
first order in our treatment we find
\begin{align}
J_{s}(T)\equiv J\kappa_{\mathrm{MF}}(T).
\end{align}
For small $T$, longitudinal fluctuation are practically frozen and
$\lim_{T\to 0}J_{s}(T)=J$.  At larger temperatures
$\kappa_{\mathrm{MF}}(T)$ decreases since longitudinal fluctuations
reduce the stiffness.  Finally, $J_{s}(T)$ vanishes at a finite
temperature value $T_{\mathrm{MF}}>T_\text{BKT}$. The mean-field
critical temperature $T_{\mathrm{MF}}$ is given by
$T_{\mathrm{MF}}=2J$ ($T_{\mathrm{MF}}=dJ$ for a hypercubic lattice in
$d$ dimensions).  To obtain this value of $T_{\mathrm{MF}}$ one has to
fix $\mu=0$.  This choice turns out to be a reasonable one, since one
finds for small $T$
\begin{align}
\frac{J_{s}(T)}{J} = 1-\frac{T}{2T_{\mathrm{MF}}}+\cdots=
1-\frac{T}{4J}+\cdots,
\label{expansion}
\end{align}
in agreement with the results of the self-consistent harmonic
approximation \cite{Pires1994}, which predicts $J_s(T)/J=1-T/zJ$ for a
model with $z$ nearest neighbors at small $T$.  In $1d$ this agrees
also with the exact low-temperature result \cite{Stanley1969}; see as
well the discussion in \cite{Benfatto2012} on the low temperature
behaviour of $J_s(T)/J$. We also mention that Monte Carlo simulations
\cite{private_comm} confirm that for low temperature one has that the
slope $\partial J_s/\partial T$ for $T \to 0$ is $1/4$, as given in
\eqref{expansion}.

In order to go beyond the saddle-point approximation, it is necessary
to explicitly solve the flow Eq.~\eqref{eq:Uflow}. Then the
expectation value $\kappa_*$ for the field $\rho$ is defined by the
minimum
\begin{align}
\frac{\partial U_{k\to0}(\rho)}{\partial\rho}\Bigl{|}_{\kappa_{*}}=0,
\end{align}
and the phase stiffness is given by \eqref{kappa_star}.

\begin{figure}[t!]
\includegraphics[width=0.45\textwidth]{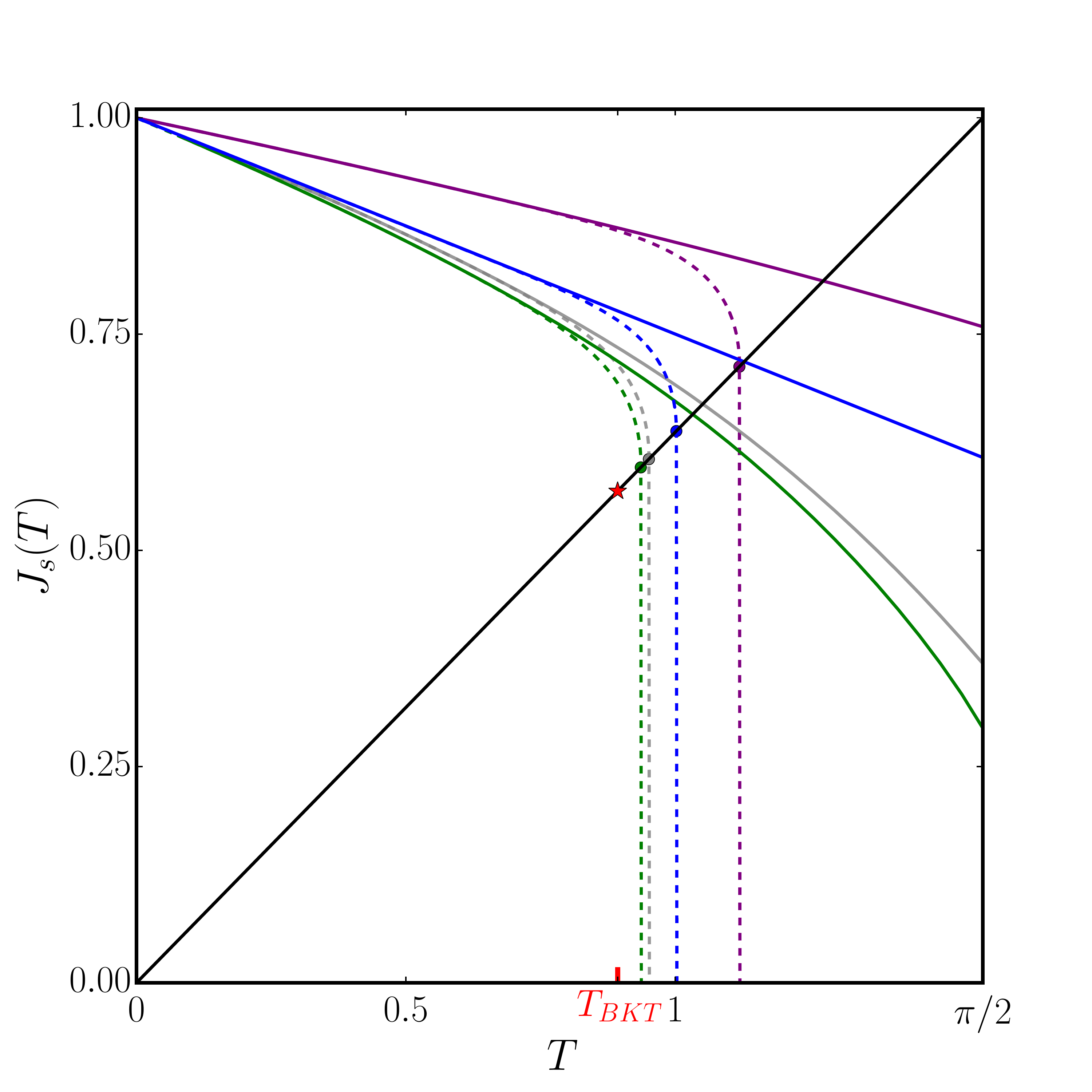}
\caption{Superfluid stiffness $J_s$ in units of $J$ as a function of
  the temperature for the $XY$ model.  The purple lines represent the
  $|\varphi|^4$ model with initial condition \eqref{EqB26} for the
  potential and $\mu=Jd$, as detailed in the case a of the main
  text. The blue lines are the case b, the gray lines the case c and
  the green lines the case d.  Solid and dashed lines represent
  respectively the results without and with the inclusion of vortex
  excitations.}
\label{Fig6}
\end{figure}

Our results are summarized in Fig.~\ref{Fig6} which shows the
temperature dependence of the spin stiffness $J_{s}(T)$.  In this
figure the solid lines correspond to the results generated by
amplitude fluctuations using Eq.~\eqref{kappa_star}, but without
considering the vortex fluctuations.  The different lines correspond
to different approximations discussed in the following.  The dashed
lines represent the vortex renormalized stiffness and are obtained by
considering the effect of vortex fluctuations via the perturbative SG
Eqs. \eqref{Eq23} and \eqref{Eq24} with initial conditions
\eqref{Eq25} and \eqref{Eq26} after performing the RG for the
amplitude modes.  Without vortex fluctuations the BKT temperature is
simply obtained by the intersection of $J_{s}(T)$ with
$\frac{2 T}{\pi}$. From top to bottom of Fig.~\ref{Fig6} we have:
\begin{enumerate}[label=\alph*.]
\item FRG, initial condition
  \eqref{EqB26}, $\mu=Jd$ (purple lines):\\
  $T_\text{BKT}/J=1.19\pm0.02$. 
\item Low-temperature expansion \eqref{expansion} (blue lines):\\
  $T_\text{BKT}/J=1.00\pm0.02$.
\item Mean-field estimate \eqref{eq:meanfield} of $J_s(T)$, $\mu=0$, (gray
  lines):\\
  $T_\text{BKT}/J=0.96\pm0.02$.
\item FRG, initial condition
  \eqref{EqB26}, $\mu=0$ (green lines):\\
  $T_\text{BKT}/J=0.94\pm0.02$.
\end{enumerate}
In the figure we also plot for comparison the Monte Carlo result
$T_\text{BKT}/J=0.893$ (red star).

A remark is in order here: when mapping the $XY$ model onto the
two-component $|\varphi|^{4}$ lattice field theory in section
\ref{sec:mapping}, we underlined that the mapping is exact and the
results should be $\mu$ independent as long as $\mu\geq Jd$.  However,
as discussed above, our FRG flow equation for the action
\eqref{eq:SAP} is applied to the $XY$ model by modifying only the
initial condition for the bare potential.  This procedure is
incomplete since the lattice field theory equivalent of the $XY$ model
has the lattice dispersion \eqref{EqB24} rather than the continuous
one \eqref{EqB25}.  Therefore, the application of the FRG flow with
continuous dispersion \eqref{EqB25} and $\mu=Jd$ (purple line in
Fig.~\ref{Fig6}) is a rather crude approximation and does not agree
with the low-temperature expansion (blue line).

Moreover, approximating the lattice dispersion with a continuous
dispersion introduces a $\mu$ dependence in our result.  We can
exploit this and fix $\mu=0$ to approach the exact low-temperature
asymptotics.  While such a value of $\mu$ would not be allowed in the
lattice theory with dispersion \eqref{EqB24}, it is permitted in the
continuous case. The resulting green solid line in Fig.~\ref{Fig6}
shows a consistent improvement over the low-temperature expansion
(blue line).

Since the effect 
of the amplitude fluctutations in the continuous $|\varphi|^4$ model 
with effective potential \eqref{EqB26} is rather small, we expect that 
analytic results for the superfluid stiffness obtained from the saddle point 
solution follow very closely the exact results in all the range 
of the temperature between zero and $T_\text{BKT}$. This can be made quantitative 
by observing that one could obtain very good results 
(plotted as gray lines in Fig.~\ref{Fig6}) by solving the following mean-field 
equation for the superfluid stiffness $J_s(T)/J$:
\begin{align}
\label{eq:meanfield}
J_s(T)= J \, \frac{I_1\left( 4\beta J_s(T) \right)}
{I_0\left( 4\beta J_s(T) \right)}
\end{align}
(which is the solid gray line), and then use it as initial condition
in the perturbative SG Eqs.~\eqref{Eq23}--\eqref{Eq24}. The procedure
gives the dashed gray line and $T_\text{BKT}=0.96\pm0.02$, worse than
the value \eqref{TBKT_anal} we find using $\mu=0$, but again
reasonably good.

In conclusion, our most accurate results come from the
  nonperturbative evaluation of the FRG flow for the amplitude mode
  combined with the perturbative SG flow for the phase:
\begin{align}
  \frac{T_\text{BKT(FRG)}}{J}=0.94 \pm 0.02,
\label{TBKT_anal}
\end{align}
in good agreement with the expected result for the XY model
$T_\text{BKT} \simeq 0.893 J$ obtained by MC simulations
\cite{Gupta1988, Schultka1994, Olsson1995, Hasenbusch2005,
  Komura2012}.  Note that this very good agreement for the critical
temperature has been obtained by matching with the appropriate choice
of $\mu$ the low-temperature behaviour of the superfluid stiffness.

\section{Conclusions}
\label{sec:concl}

The topological phase transition in two-dimensional spin models with
continuous symmetry as explained by the Berezinskii, Kosterlitz and
Thouless (BKT) theory is a celebrated result.  Our aim in this paper
has been to set up and implement a renormalization group framework for
the BKT universality class to quantitatively determine nonuniversal
properties such as the temperature dependence of the superfluid
fraction, the critical chemical potential and the transition
temperature, given their relevance in $2d$ realizations of BKT physics
and in current experiments.

After discussing the role of the parametrization of the field in
functional RG approaches to $2d$ BKT phase transitions, we argue that
the amplitude-phase (AP) Madelung representation of the field is the
natural choice to study the contribution of longitudinal spin
fluctuations to nonuniversal quantities and we show that amplitude
fluctuations are gapped at the critical point.  With the AP
parametrization we have been able to study the RG flow directly in the
relevant degrees of freedom: amplitude (density) fluctuations,
longitudinal spin waves, and vortex excitations, and we discuss their
mutual interplay.

As a preliminary step, we have derived an explicit mapping from the
$2d$ lattice $XY$ model to a continuum $|\varphi|^4$ field theory.
While in three and higher dimensions this continuum limit is
straightforward, in two dimensions the mapping depends, qualitatively
and quantitatively, on nonuniversal ultraviolet details of the initial
model. As a result, we have mapped the original $XY$ coupling $J$ to
the initial superfluid stiffness $\rho$ and interaction $\lambda$ at
cutoff scale $\Lambda$ of the corresponding $|\varphi|^4$ model.
Therefore, the RG equations are the same and only the initial
conditions differ to characterize the $XY$ and $|\varphi|^4$ models,
so that they can be treated within the same formalism on equal
footing.

We then proceeded to write the action in the amplitude and phase
degrees of freedom and we have shown that amplitude excitations
are gapped, such that the BKT behavior is correctly recovered as a
transition and not as a crossover at large distances.  This result is
based on the explicit subtraction in the functional RG equations of
the Gaussian energy.  While this is mainly a technical point, we think
it is an interesting one since (i) in many other applications
such contributions do not have any physical effect in the
determination of the critical properties of $O(N)$ models, and
(ii) the AP representation provides a straightforward way to
show this effect.

Our FRG procedure is then based on two steps: we first perform FRG on
the amplitude part $S_A$ of the action \eqref{eq:Srho}. We then
insert the obtained stiffness into the phase part of the action,
which is given by the spin-wave action \eqref{eq:Sphase} with the
phase crucially considered as a periodic variable. This allows us to
correctly take into account the compact nature of the phase variable
and to use the results of the sine-Gordon model.

The combination of the nonperturbative functional RG analysis of the
amplitude part of the action with the perturbative flow for the
sine-Gordon model is already sufficient to give rather good results
for nonuniversal and universal quantities.  In particular, we
determined the critical chemical potential for the $|\varphi|^4$ model
and the nontrivial universal parameters $\kappa'$ and $\theta_0$
defined in Eqs.~\eqref{kappa_MC} and \eqref{theta0_MC}.  Our results
for these two parameters are $\kappa_\text{(FRG)}'=0.67\pm 0.07$ and
$\theta_{0\text{(FRG)}}=1.033\pm 0.032$, which should be compared with
the Monte Carlo results $\kappa'=0.61 \pm 0.01$ and
$\theta_0=1.068 \pm 0.01$ \cite{Prokofev2002}.  For the $XY$ model we
obtained the temperature dependence of the stiffness $J_s(T)$, which
receives nonuniversal corrections from amplitude fluctuations. It
reproduces the exact low-temperature limit and predicts the critical
temperature with an error of $\approx 5\%$.

In conclusion, our findings confirm that amplitude fluctuations only
result in a finite renormalization of the stiffness and do not
completely deplete the superfluid fraction.  We also find, without
{\em a priori} assumptions, that amplitude fluctuations are frozen for
the $|\varphi|^4$ model and yield effectively a phase-only model of
spin-wave and vortex excitations. Finally, we proved that the combined
use of the functional RG for the amplitude modes and of perturbative
results for the sine-Gordon model allows one to quantify the effect of
vortex excitations at finite temperature, which depends on the value
of the vortex core energy and yields a further lowering of $T_c$
\cite{Benfatto2012}. Results for several universal and nonuniversal
quantities are presented, with a very good agreement with known
results.

To further improve the results obtained for both the $|\varphi|^4$ and
the $XY$ models, one can include nonperturbative effects in the
sine-Gordon part of the RG flow.  To this end, one should compute the
anomalous dimension $\eta$ in the nonperturbative RG flow of the
sine-Gordon model.  Moreover, for the $XY$ model, one should include
lattice effects \cite{Machado2010} which are beyond the
scope of this paper.  The study of lattice effects leads in a natural
way to generalized sine-Gordon models, which we think is promising for
future work.  Although the obtained results are rather good, we think
that the nonperturbative treatment of the SG part of the action, and
of the lattice effects for the $XY$ model, may lead to further
improvements that are worthwhile to estimate.

This work can provide a basis for future efforts to derive a
generalized sine-Gordon model which comprehensively includes amplitude
fluctuations on equal footing with phase fluctuations, and not as an
initial condition from a previous RG step, as we did in this paper.
In this way one should be able to describe also the feedback of vortex
excitations onto amplitude fluctuations.  We think that it would be
interesting to extend the results of this work to $2d$ quantum systems
in order to quantitatively determine $T_c$ as a function of
interaction strength in ultracold Bose \cite{Hadzibabic2006,
  Schweikhard2007} and Fermi gases \cite{Bauer2014, Ries2015,
  Murthy2015, Madeira2017, Murthy2017} and for out-of-equilibrium
situations \cite{Chu2001, He2017, Schweigler2017}.

\textit{Acknowledgements.}  The authors wish to thank L.~Benfatto,
C.~Castellani, N.~Dupuis, G.~Gori, Z.~Gulacsi, H.~Kn\"orrer,
J.~Lorenzana, and I.~Maccari for insightful discussions.  We also
thank M.~Hasenbusch and N.~Prokofev for useful correspondence.  This
work is part of and supported by the DFG Collaborative Research Centre
``SFB 1225 (ISOQUANT)''.  T.E. thanks the Erwin-Schr\"odinger
Institute in Vienna for hospitality during the initial stages of this
work.  This work was supported by the J\'anos Bolyai Research
Scholarship of the Hungarian Academy of Sciences.  A.T. and
I.N. acknowledge support from Progetto Premiale ABNANOTECH. N.D. and
A.T. thank the Institut Henri Poincar\'e--Centre Emile Borel for
hospitality, where the final part of this work was done during the
trimester ``Stochastic Dynamics Out of Equilibrium''. \\

\textit{Note added.} After the submission of this paper, an FRG
treatment of the $XY$ model by J. Krieg and P. Kopietz
\cite{Krieg2017} appeared on arXiv.  Within the Coulomb gas
representation, these authors include amplitude fluctuations
perturbatively and find a true line of fixed points, confirming the
importance of explicitly using vortex degrees of freedom.  The main
difference is that our approach treats amplitude fluctuations
nonperturbatively, while \cite{Krieg2017} includes lattice effects
explicitly \cite{Machado2010}.

\appendix

\section{Spin-wave approximation}
\label{AppendixA}

The expression for the magnetization is given by
\begin{align}
M_{i}=\left\langle \frac{e^{i\theta_{i}}}{2}\right\rangle+\left\langle \frac{e^{-i\theta_{i}}}{2}\right\rangle,
\end{align}
while the expression of the spin-spin correlation function on the
lattice is
\begin{align}
G_{ij}=\langle \cos(\theta_{i}-\theta_{j})\rangle.
\end{align}
Both are conveniently rewritten in continuous notation as
\begin{align}
F(x)&=\int\mathcal{D}\theta\,e^{\int\left[-\frac{K}{2}\left(\nabla\theta\right)^{2}+J(x')\theta(x')\right]d^{d}x'}
\end{align}
with $J(x')=i\delta(x')$ and $J(x')=i\delta(x-x')-i\delta(x')$, where
the two expressions are valid respectively for the magnetization and
the two-point correlation function. The integral in latter expression
yields
\begin{align}
\label{EqA3}
F(x)&=e^{\int{\left[\frac{1}{2K}J(x')\mathcal{G}(x'-y')J(y')\right]d^{d}x'\,d^{d}y'}}\nonumber\\
&=\begin{cases}
&M(x)=e^{-\frac{1}{K}\mathcal{G}(0)}\\
&G(x)=e^{\frac{1}{K}\left[\mathcal{G}(x)-\mathcal{G}(0)\right]}\end{cases}
\end{align}
where 
\begin{align}
\mathcal{G}(x)=\int{\frac{d^{d}q}{(2\pi)^{d}}\frac{e^{-i\vec q\cdot \vec x}}{q^{2}}}
\end{align}
(the $x=0$ case must be evaluated separately in a finite volume and in
the thermodynamic limit). In a finite system of size $L$ we obtain in
$d=2$
\begin{align}
G_{L}(0)=\frac{1}{2\pi}\log\left(\frac{L}{a}\right),
\end{align}
leading to a vanishing magnetization in the $2d$ system in the
thermodynamic limit.

In order to evaluate $\mathcal{G}(x)$ it is convenient to perform the
computation directly in the thermodynamic limit.  We first consider a
general dimension $d$ and then compute the $d\to 2$ limit. One has
then
\begin{align}
\mathcal{G}(0)=\frac{s_{d}\pi^{d-2}}{d-2}a^{2-d},
\end{align}
where $s_{d}$ is the surface of the $d$-dimensional unit sphere
divided by $(2\pi)^{d}$. The finite $x$ expression can be obtained in
the continuum limit $a\to 0$ as
\begin{align}
\mathcal{G}(x)=\frac{s_dx^{2-d}}{d-2},
\end{align}
and one obtains
\begin{align}
\lim_{d\to 2}\left[\mathcal{G}(x)-\mathcal{G}(0)\right]=-\frac{1}{2\pi}\log\left(\frac{\pi x}{a}\right).
\end{align}

\section{Flow equations for the amplitude and phase scheme}
\label{AppendixC}

In order to derive the FRG flow equations, we project the Wetterich
equation\cite{Wetterich1993} onto the theory space defined by the
effective action ansatz \eqref{eq:Gammak} to obtain \cite{Berges2002}
\begin{align}
&\partial_{t}U_{k}(\rho)=\nonumber\\
&\frac{1}{2}\int\frac{d^{d}q}{(2\pi)^d}\left[\frac{\partial_{t}R^{(\theta)}_{k}(q)}{\rho\,q^{2}+R^{(\theta)}_{k}(q)}+\frac{\partial_{t}R^{(\rho)}_{k}(q)}{(4\rho)^{-1}q^{2}+U_{k}^{(2)}(\rho)+R^{(\rho)}_{k}(q)}\right].
\label{EqD1}
\end{align}
We choose both amplitude and phase regulators $R^{(\ell)}$, with
$\ell=\rho,\theta$, of the form
\begin{align}
\label{EqC2}
R_{k}^{(\ell)}(q)=\alpha_{\ell}(k^{2}-q^{2})\theta(k^{2}-q^{2}),
\end{align}
where $\alpha_{\ell}$ is a dimensional coefficient necessary to have
the correct scaling dimension of the regulator terms.  The scale
derivative of the regulator is then
\begin{align}
\partial_{t}R_{k}^{(\ell)}(q)=-(2\,\alpha_{\ell}-\partial_t\alpha_{\ell}) 
k^{2}\theta(k^{2}-q^{2}).
\end{align}
These $\theta$ functions in the numerator of the integral in
Eq.~\eqref{EqD1} constrain the momenta to $q^{2}\in[0,k^{2}]$, where
the regulator $\theta$ functions in the denominators are always unity.
We are then left with the calculation of two integrals of the type (in
$d=2$)
\begin{align}
\frac{1}{2\pi}\int_{0}^{k^{2}}k^{2}\left(a q^{2}+b\right)^{-1}qdq,
\end{align}
where $a$ and $b$ are two $q$-independent constants.
It is convenient to define the rescaled variable $x=q^{2}/k^{2}$ leading to
\begin{align}
\frac{k^{2}}{4\pi}\int_{0}^{1} \left(a x+\frac{b}{k^{2}}\right)^{-1}dx
  =\frac{k^{2}}{4\pi a}\log\left(1+ak^2/b\right).
\end{align}
Substituing $a$ and $b$ with the coefficients of the integrals in
Eq.~\eqref{EqD1}, one obtains the full potential flow equation
\begin{multline}
\partial_{t}U_{k}(\rho)=-\frac{k^2}{4 \pi }\left(\frac{\alpha_{\theta}
  \log \left(\rho/\alpha_{\theta}\right)}{\rho -
  \alpha_{\theta}}\right.\\
  +\left.\frac{4\alpha_{\rho}  \rho  \log \left(1+\frac{4\alpha_\rho  \rho -1}{4 \rho  U_k''(\rho
   )/k^2+1}\right)}{4\alpha_{\rho}  \rho -1}\right),
\label{EqB6}
\end{multline}
where we have used the fact that $\partial_{t}\alpha_{\ell}=0$ in two
dimensions. The flow for Gaussian theories $U''_{k}(\rho)=0$ is simply
\begin{multline}
\partial_{t}U_{k}(\rho)=-\frac{k^2}{4 \pi }\left(\frac{\alpha_{\theta}
  \log \left(\rho/\alpha_{\theta}\right)}{\rho -
  \alpha_{\theta}}\right.\\
  +\left.\frac{4\alpha_{\rho}  \rho  \log \left(4\alpha_\rho\rho\right)}{4\alpha_{\rho}  \rho -1}\right).
\label{EqB7}
\end{multline}
According to the discussion in the text the flow for Gaussian theories
must vanish, thus, in order to enforce this condition, we simply
subtract the r.h.s. of Eq.\,\eqref{EqB7} from the r.h.s. of
Eq.\,\eqref{EqB6}.  The latter procedure finally produces
Eq.\,\eqref{eq:Uflow} in the text.

This equation is solved numerically for the full
potential function to produce the numerical results shown in
Section~\ref{sec:res}.  Nevertheless, in order to gain a qualitative
understanding of the flow, it is useful to employ a second-order
Taylor expansion around the running potential minimum
\begin{align}
\label{EqC5}
U_{k}(\rho)=\frac{\lambda_{k}}{2}(\rho-\kappa_{k})^{2},
\end{align}
which leads to the following flowing RG couplings:
\begin{align}
\partial_{t}\kappa_{k}&=-\frac{\partial_{t}U_{k}^{(1)}(\kappa_{k})}{U_k^{(2)}(\kappa_{k})},\label{EqC6}\\
\partial_{t}\lambda_{k}&=\partial_{t}U_{k}^{(2)}(\kappa_{k})+U_{k}^{(3)}(\kappa_{k})\partial_{t}\kappa_{k}\label{EqC7}.
\end{align}
The general flow equation \eqref{EqB6} contains two free parameters
$\alpha_{\theta,\rho}$, which are dimensionless in $d=2$.  The phase
diagram in Fig.~\ref{Fig3} has been obtained with
$\alpha_{\theta}=\kappa_{k}$ and
$\alpha_\rho=1/\left(4\kappa_{k}\right)$ in order to simplify the flow
equations, but different choices of these parameters give equivalent
results.

\bibliographystyle{apsrev_cm}
\bibliography{BKT}
\end{document}